\documentclass[sn-mathphys]{sn-jnl}

\usepackage{bm}
\usepackage{multirow}
\usepackage{colortbl}
\usepackage[utf8]{inputenc}
\usepackage{graphicx}
\usepackage{epstopdf}
\usepackage{csquotes}
\usepackage{amsmath}
\MakeOuterQuote{"}
\usepackage{mathtools}
\DeclarePairedDelimiter\abs{\lvert}{\rvert}%
\DeclarePairedDelimiter\norm{\lVert}{\rVert}%

\jyear{2021}%

\theoremstyle{thmstyleone}%
\theoremstyle{thmstyletwo}%

\theoremstyle{thmstylethree}%

\begin{document}

\title[Optimal parameters for time-stepping global stability analysis]{Optimal computational parameters for maximum accuracy and minimum cost of Arnoldi-based time-stepping methods for flow global stability analysis}


\author*[1]{\fnm{Marlon} \sur{Sproesser Mathias}}\email{marlonsmathias@gmail.com}

\author[1]{\fnm{Marcello Augusto} \sur{Faraco de Medeiros}}\email{marcello@sc.usp.br}

\affil*[1]{\orgdiv{São Carlos School of Engineering}, \orgname{University of São Paulo}, \orgaddress{\street{Av. João Dagnone}, \city{São Carlos}, \postcode{13563-120}, \state{São Paulo}, \country{Brazil}}}


\abstract{Global instability analysis of flows is often performed via time-stepping methods, based on the Arnoldi algorithm. When setting up these methods, several computational parameters must be chosen, which affect intrinsic errors of the procedure, such as the truncation errors, the discretization error of the flow solver, the error associated with the nonlinear terms of the Navier-Stokes equations and the error associated with the limited size of the approximation of the Jacobian matrix. This paper develops theoretical equations for the estimation of optimal balance between accuracy and cost for each case. The 2D open cavity flow is used both for explaining the effect of the parameters on the accuracy and the cost of the solution, and for verifying the quality of the predictions. The equations demonstrate the impact of each parameter on the quality of the solution. For example, if higher order methods are used for approaching a Fréchet derivative in the procedure, it is shown that the solution deteriorates more rapidly for larger grids or less accurate flow solvers. On the other hand, lower order approximations are more sensitive to the initial disturbance magnitude. Nevertheless, for accurate flow solvers and moderate grid dimensions, first order Fréchet derivative approximation with optimal computational parameters can provide 5 decimal place accurate eigenvalues. It is further shown that optimal parameters based on accuracy tend to also lead to the most cost-effective solution. The predictive equations, guidelines and conclusions are general, and, in principle, applicable to any flow, including 3D ones.}

\keywords{Time-stepping global stability, Linear stability theory, Compressible flow, Open cavity flow}



\maketitle

\section{Introduction}

Methods for computing the global stability of flows have been used on several applications in the past decades \citep{Theofilis2011}. Many implementations involve building the Jacobian matrix of the flow and solving its eigenproblem. However, starting with Eriksson and Rizzi\cite{Eriksson1985}, who worked with the Euler equations, the alternative, time-stepping method was developed and became popular. Edwards et al.\cite{Edwards1994} and Chiba\cite{Chiba1998} used this method for the Navier-Stokes equations and Tezuka and Suzuki\cite{Tezuka2006}, to solve a 3D global stability case for the first time. These methods do not require the Jacobian matrix explicitly, but, instead, require a flow solver.

One reason for the popularity of time-stepping methods is that the flow solver can be treated as a black box, which is simply incorporated into the algorithm, greatly reducing code implementation and verification efforts. Gómez et al.\cite{Gomez2014} and Liu et al.\cite{Liu2016} demonstrated the great flexibility of this method by coupling it to the well-known OpenFOAM CFD software. Time-stepping methods have also been successfully used to solve the adjoint eigenproblem for flow sensitivity analysis \citep{Brynjell-Rahkola2017} and for transient growth analysis \citep{Barkley2008}. Mack and Schmid \cite{Mack2010} implement a time-stepping solver and use Cayley transformations to accelerate the convergence of the algorithm in specific parts of the spectrum. More recently, Negi et al.\cite{Negi2020} have employed this method for fluid-structure interaction problems. A review of similar methods and its applications can be found in Loiseau et al.\cite{Loiseau2019}. However, if a very large number of eigenvalues is required, the matrix-forming methods are usually more efficient \citep{Theofilis2011}.

The widely used time-stepping approach discussed in this work is based on the Arnoldi method \citep{Arnoldi1951} for solving eigenproblems. Instead of needing the flow's Jacobian matrix, this method only requires the ability to obtain the product of such matrix by an arbitrary vector. The governing equations are rearranged in a way that the result of this product can be obtained from a run of a flow solver.

In this paper we investigate parameters that influence the accuracy and cost of this method and develop equations that provide optimal parameter values for its efficient use. We will use the classical open cavity flow as an example. Even though the specific optimal parameters will vary according to, for example, the flow studied, flow solver used and the computational resources available, the trend and sensibility of the method's performance with respect to each parameter, the  guidelines , the equations for estimating optimal values and the conclusions are general.

The paper focuses on four choices that need to be made when setting up this algorithm: the disturbance magnitude; the Fréchet derivative accuracy order; flow simulation time for each Arnoldi iteration; and the size of the Krylov span. Section~\ref{sec:methods} covers the methods employed in our global stability analysis and explains these parameters and where each of them is needed. As will be explained, even though all these parameters affect accuracy in complicated ways, the optimal value of disturbance magnitude is mostly independent on the integration time and on the size of the Krylov span. In turn, the optimal integration time is also independent on the Krylov span dimension and is only weakly dependent on the disturbance magnitude.

The first two choices are discussed in section~\ref{sec:epsilon} and are associated to the trade-off between Fréchet truncation errors and flow solver discretization errors. Section~\ref{sec:tau} discusses the effect of the integration time. Section~\ref{sec:runTime} analyses the trade-off between the Krylov span size and the integration time of each Arnoldi iteration. Section~\ref{sec:eigfunctions} covers the convergence of eigenmodes as well as the contribution of each Arnoldi iteration to the overall result. The last section summarizes the conclusion.

Goldhirsch et al.\cite{Goldhirsch1987} developed a similar method for obtaining the leading eigenvalues of a large matrix by converting it into a time-dependent differential equation and using this new system to select only the leading modes. They analyzed some sources of errors and part of the current study extends their work. We are concerned with the accuracy in retrieving flow modes. The important topic of the emergence of spurious modes owing to ill-posed boundary conditions is outside our scope; this is discussed by Ohmichi and Suzuki\cite{Ohmichi2016}. Our results do not apply to matrix forming methods, whose accuracy and efficiency are analyzed by Gennaro et al.\cite{Gennaro2013}.

\section{Methods}
\label{sec:methods}

\subsection{Linear stability computations}
\label{sec:linStab}

A flow evolution can be expressed as

\begin{equation}
	\frac{d {U}}{d t} = {f}({U}) \label{eq:NavierStokes} \ ,
\end{equation}

\noindent where ${U}$ is a vector that contains all flow variables for each node of a mesh. The flow ${U}$ is given by the base flow plus a disturbance: ${U} = {U}_0 + {\tilde{u}}$. The known base flow satisfies the steady Navier-Stokes equations, and the disturbance is small. $A$ is the Jacobian matrix

\begin{equation}
	{A} = \frac{\partial {f}({U_0})}{\partial {U}} \label{eq:Jacobian} \ ,
\end{equation}

\noindent hence, after linearizing equation \ref{eq:NavierStokes} around $U_0$, we obtain

\begin{equation}
	\frac{d \tilde{u}}{d t} = A \tilde{u} \ .
\end{equation}

For an initial disturbance $\tilde{u}_0$, the solution ${\tilde{u}}$ is, formally,

\begin{equation}
	\tilde{u} = e^{t {A}} \tilde{u}_0 \label{eq:exactLinearSolution} \ .
\end{equation}

By computing the eigenvalues ${S}$ and eigenvectors ${V}$ of ${A}$, equation~\ref{eq:exactLinearSolution} can be rewritten as

\begin{equation}
	\tilde{u} = {V} e^{t {S}} {V}^{-1} \tilde{u}_0 \label{eq:exactLinearSolutionDiag} \ .
\end{equation}

\noindent Each column ${\phi_A}$ of ${V}$ corresponds to a flow mode and is associated to a value $\sigma_A$ in $S$, which gives the mode's amplification rate and oscillation frequency as its real and imaginary parts, respectively. Positive amplification rates indicate instability.

The Jacobian matrix is $N\times N$, where $N$ is the number of nodes in the domain multiplied by the number of dependent variables, namely, 4 for 2D flows and 5 for 3D ones, when using the compressible Navier-Stokes equations; hence the interest in methods that avoid the solution of the full Jacobian matrix. In the approach used here \citep{Eriksson1985, Chiba1998, Tezuka2006, Gomez2014}, it is convenient to define a matrix

\begin{equation}
	B=e^{\tau A} \label{eq:Bdef} \ .
\end{equation}

The concept behind the eigenvalue solver method used here is that by recursively multiplying a matrix ${B}$ by a vector, the eigenvector corresponding to the eigenvalue with the greatest absolute value is selected. Using this idea, it is possible to construct a set of $n$ linearly independent vectors that approximate the eigenvectors of the $n$ largest eigenvalues of ${B}$, which correspond to the leading eigenvalues of ${A}$. If care is taken, the approximation can be very good. The method is implemented via the Arnoldi algorithm \citep{Arnoldi1951} and belongs to the class of orthogonal projection techniques for approximate solutions of linear systems \citep{Saad2003}.

The Arnoldi algorithm starts with a $\ell^2$ normalized vector ${\zeta}_1$, which will be discussed in the next section. At each iteration $k$, the Arnoldi algorithm takes a normalized vector ${\zeta}_k$, and calculates the product $B\zeta_k$.  Next, the dot products of ${\zeta}_j$ ($j=1\dots k$) with ${B}{\zeta}_k$ are stored in positions $h_{j,k}$ of matrix $H$. A vector $\tilde{u}_{k+1}$ is given by the Gram-Schmidt orthonormalization of ${B}{\zeta}_{k}$ with respect to all previous ${\zeta}_j$ ($j = 1 \dots k$). The norm of this vector is stored in $h_{k+1,k}$. The disturbance distribution vector for the following iteration, ${\zeta}_{k+1}$, is obtained by normalizing $\tilde{u}_{k+1}$. The matrix $H$ is of the Hessemberg form, and is constructed so that \citep{Tezuka2006}

\begin{equation}
	{B} \left[{\zeta}_1, {\zeta}_2, \dots, {\zeta}_M\right] = \left[{\zeta}_1, {\zeta}_2, \dots, {\zeta}_M\right] \left[\begin{array}{ccccc}
		h_{1,1} & \dots &  & \dots & h_{1,M} \\
		h_{2,1} & \dots &  &  & \vdots \\
		0 & \ddots &  &  &  \\
		\vdots & \ddots & \ddots & \vdots & \vdots \\
		0 & \dots & 0 & h_{M,M-1} & h_{M,M}
	\end{array} \right] \ .
	\label{eq:Hessenberg}
\end{equation}

After enough Arnoldi iterations, the eigenvalues of ${B}$, ${\sigma}_B$, with the largest modulus are approximated by the eigenvalues of ${H}$. In turn, the eigenvalues of ${A}$ are ${\sigma}_A=\log{({\sigma}_B)}/t \approx \log{({\sigma}_H)}/t$. The respective eigenvectors are given by $\phi_A=\phi_B \approx [ \zeta_1 \dots \zeta_M ] \phi_H$. The $H$ matrix is orders of magnitude smaller than the original Jacobian matrix of the flow problem, which massively reduces the computational resources needed. In calculating the eigenvalues of $A$ from those of $B$, the procedure is ambiguous because the complex logarithm function is multivalued. The imaginary part, which relates to the frequency, has an undetermined contribution of $2\pi n/\tau$, with $n$ integer. Its value must be verified by, for example, observing the mode oscillation in the flow simulation. This ambiguity of the imaginary part does not affect the accuracy, and often it is the real part that is most vulnerable to discretization errors.

Finally, to multiply $B$ by $\zeta_k$, define $F$ as

\begin{equation}
	F(U) = U(0) + \int_0^\tau f (U) dt \ ,
	\label{eq:taylor1}
\end{equation}

\noindent where $\tau$ is hereinafter referred to as integration time. The flow $U(t)$ is given by the sum $U_0 + \tilde{u}(t)$, in which $\tilde{u}(0) = \tilde{u}_0$ and $\tilde{u}(\tau)$ is given by equation~\ref{eq:exactLinearSolution}. Substituting these in equation~\ref{eq:taylor1} and $e^{\tau A}$ by $B$ shows that F can provide an approximation for the linear model as:

\begin{equation}
	F(U) \approx U_0 + B \tilde{u}_0 \ .
	\label{eq:jacobian1}
\end{equation}

\noindent Hence, $B$ is the Jacobian of $F$:

\begin{equation}
	\frac{\partial F(U)}{\partial \tilde{u_0}} = B \ .
	\label{eq:jacobian4}
\end{equation}

\noindent Using a Taylor series expansion, we can write
\begin{equation}
	F(U_0 + \varepsilon_0 \zeta) = F(U_0) + \varepsilon_0 B \zeta + \overrightarrow{\mathcal{O}}(\norm{\varepsilon_0 \zeta}^2)\ ,
	\label{eq:taylor3}
\end{equation}

\noindent where $\varepsilon_0$ is a small scalar. Keeping in mind that $\zeta$ has unitary norm, this leads to

\begin{equation}
	{B} {\zeta} = \frac{1}{\varepsilon_0} \left[ {F}\left({U}_0 + \varepsilon_0 {\zeta}\right) - {F}\left({U}_0\right) \right] + \overrightarrow{\mathcal{O}}(\varepsilon_0) \ .
	\label{eq:frechet1}
\end{equation}

\noindent $B \zeta$ is the directional derivative of $F$ in the $\zeta$ direction, known as Fréchet derivative. Equation~\ref {eq:frechet1} provides a first-order accurate approximation for $B \zeta$.

A second-order approximation is given by:

\begin{equation}
	{B} {\zeta} = \frac{1}{2\varepsilon_0} \left[ {F}\left({U}_0 + \varepsilon_0 {\zeta}\right) - {F}\left({U}_0  - \varepsilon_0 {\zeta}\right) \right] + \overrightarrow{\mathcal{O}}(\varepsilon_0^2) \ ;
	\label{eq:frechet2}
\end{equation}

\noindent and a fourth-order:

\begin{multline}
	{B} {\zeta} = \frac{1}{12\varepsilon_0} [ -{F}\left({U}_0 + 2\varepsilon_0 {\zeta}\right) + 8{F}\left({U}_0 + \varepsilon_0 {\zeta}\right) \\
	-8{F}\left({U}_0  - \varepsilon_0 {\zeta}\right) + {F}\left({U}_0 -2\varepsilon_0 {\zeta}\right) ] + \overrightarrow{\mathcal{O}}(\varepsilon_0^4) \ .
	\label{eq:frechet4}
\end{multline}

\noindent Most of the literature chose the second-order approximation \citep{Chiba1998,Tezuka2006,Gomez2015a}, while, for example, Eriksson and Rizzi\cite{Eriksson1985} used the fourth-order variant. In this paper this choice is analyzed. The values $F$ to be used in equations~\ref{eq:frechet1}~to~\ref{eq:frechet4} are obtained by integrating equation~\ref{eq:NavierStokes} numerically for a given disturbance $\tilde{u}_0$. This is carried out by the flow solver. The initial disturbance is discussed in the next section and the flow solver used is described later. Algorithm~\ref{algorithm} summarizes the full process.

Algorithm~\ref{lst:algorithm} summarizes the full process.

\begin{algorithm}
	\caption{}\label{algorithm}
	\begin{algorithmic}[1]
		\State Steady state flow ${U}_0$ from flow solver
		\State Initial disturbance ${\zeta}_1$ of norm 1
		\For{$k = 1 \dots M$}
		\State Compute ${B\zeta}_k$ using equation~\ref{eq:frechet1},~\ref{eq:frechet2},~or~\ref{eq:frechet4}
		\State $h_{j,k} \gets {\zeta}_j^T \cdot {B\zeta}_k \qquad j = 1 \dots k$
		\State $\tilde{\zeta}_{k+1} \gets {B\zeta}_k - \sum_{j=1}^{k} h_{j,k} {\zeta}_j$
		\State $h_{k+1,k} \gets \norm{\tilde{\zeta}_{k+1}}$
		\State ${\zeta}_{k+1} \gets \tilde{\zeta}_{k+1} / h_{k+1,k}$
		\EndFor
		\State Compute eigenvalues $\sigma_H$ and eigenvectors $\phi_H$ of ${H}$
		\State $\sigma_A \gets \log{(\sigma_H)}/\tau$
		\State $\phi_A \gets \{ \zeta_1 \dots \zeta_k \} \phi_H$
	\end{algorithmic}
\end{algorithm}

Two different sources cause most of the errors. One is the Arnoldi algorithm, which uses the reduced order matrix $H$ to approximate the large matrix $B$. Increasing the number of Arnoldi iterations ($M$) creates a larger Krylov span which can more accurately reproduce the true modes of the flow. However, $\tau$ also affects this approximation, and the product $\tau M$ is what governs this accuracy \citep{Goldhirsch1987}. In sections~\ref{sec:runTime}~and~\ref{sec:eigfunctions}, we present results for various values of $M$ and $\tau$ and analyze the convergence of different modes.

The other error is the inaccuracy in approximating $B \zeta_k$ in equations~\ref{eq:frechet1}~to~\ref{eq:frechet4} and can be further divided into two parts: the errors in the Fréchet derivative approximation and errors associated with the evaluation of $F$ via the flow solver. They are affected by the choices of truncation order in the Fréchet derivative, $\varepsilon_0$ and $\tau$. The error in evaluating $F$ is associated with discretization error and nonlinear terms. Naturally, linearized flow solvers are not subject to nonlinear error, such as the one used by Loiseau et al.\cite{Loiseau2014}, however, often the flow solver is not linearized. Sections~\ref{sec:epsilon}~and~\ref{sec:tau} explore these issues and make it clear that some parameters affect more than one source of error.

\subsection{Initial disturbance}

The Arnoldi method efficiency also depends on the initial vector ($\zeta_1$), as modes with a larger contribution to the initial disturbance tend to converge faster. In principle, the initial disturbance should contain all desired modes. However, Goldhirsch et al.\cite{Goldhirsch1987} argue that even if a certain mode is not contained in the initial disturbance, it is likely to appear in the result due to numerical errors.

Random distributions, such as white noise, which have a flat spectrum may be seen as an advantage. However, they excite high-frequency modes, which are soon damped out, resulting in a quick loss of disturbance energy. We used as the initial disturbance a Gaussian distribution centered close to the cavity, where most of the activity takes place. The characteristic lengths in each direction are chosen so that the disturbance approaches zero well before the boundaries of the domain.

Equations~\ref{eq:frechet1}~to~\ref{eq:frechet4} require the selection of a small scalar $\varepsilon_0$, which is the norm of the disturbance $\tilde{u_0}$ that is added to the flow. It is convenient to also define a parameter $\varepsilon$, such that $\varepsilon=\varepsilon_0/ \sqrt{N}$, where $N$ is the total number of variables in the flow solver and remains constant in our study. $\varepsilon$ is the RMS of the disturbance $\tilde{u_0}$.

The error when computing $B \zeta$ has two major contributors: the truncation error in the Fréchet derivative and the accuracy of the $F$ approximations. The former is directly linked to the norm of the disturbance ($\varepsilon_0$), while the latter is related to its RMS ($\varepsilon$).

\subsection{Computational cost}
\label{sec:compCost}

One objective of this paper is to provide guidelines for accessing efficiency in computing the modes with a desired accuracy. There are several measures of computational cost. CPU time is often used; however, this can be difficult to establish if parallel processing is used. In our case, while most parts of the computational procedures are parallelized, in some of them, notably the eigenvalue solution, the algorithms are closed, and we have limited access to the parallelization level achieved. Moreover, CPU cost is especially important with large clusters shared by many users. In our case, the code was allowed full access to 10 CPU cores of a workstation. Under these circumstances the time to a solution is an interesting measure of cost. This measure was used here and referred to as wall-time; however, as discussed at the end of the paper, the conclusions can be extrapolated to other measures of cost.

Often a large part of the computational demand comes from the flow solver runs which provides the product $B \zeta$. In our framework, this cost is proportional to the number of Arnoldi iterations $M$, yielding a computational complexity of $\mathcal{O}(M)$. Some steps involve computational complexities above $\mathcal{O}(M)$; for instance, the Gram-Schmidt orthonormalization in the Arnoldi algorithm has a complexity of $\mathcal{O}(M^2)$ and the eigenvalue solution of $H$, despite optimizations that can be used in some cases, tends to $\mathcal{O}(M^3)$ complex. There is also an overhead for transferring data between the flow solver and the global stability solver, as well as an overhead for starting up the flow solver at each Arnoldi iteration. Hence a total cost estimate of the method is given by:

\begin{equation}
	C = \left(C_\tau \: \tau + C_{It}\right) N_{call} \: M + C_{GS} \: M^2 + C_{Eig} \: M^3 + C_0 \ ,
	\label{eq:computationalCost}
\end{equation}

\noindent where $\tau$ is the integration time; $C_\tau\tau$ is the wall-time the flow solver takes to integrate until $\tau$; $C_{It}$ is the wall-time of calling the flow solver once, which includes the IO operations between the flow solver and the eigenvalue problem solver, and the flow solver startup overhead; $N_{call}$ is the number of flow solver calls to form one column in the Krylov span and depends on the order of accuracy of the Fréchet derivative; $C_{GS}$ is the wall-time of the Gram-Schmidt operation, normalized by $M^2$. $C_{Eig}$ is the wall-time of computing the eigenvalues of $H$, normalized by $M^3$; $C_0$ includes all other overheads, such as start-up and compilation time, and is fixed and negligible in the scope of this study.

Our Arnoldi algorithm implementation is in MATLAB and uses its built-in parallelization capabilities; the flow solver is written in FORTRAN and is parallelized by MPI. The cost coefficients were estimated with the aid of MATLAB's profiling tool and averaged across many executions with different parameters, in the ranges $100\le M \le 4000$ and $0.01 \le \tau \le 10$. Table~\ref{tab:computationalCostCoeff} shows these values in seconds. They are highly dependent on the implementation, on the computer and on the specific details of each flow case. Nevertheless, they provide an idea of the order of magnitude of the overall cost and how it scales with the Krylov span dimension, $M$. In our investigation, the discretization time interval of the flow solver was fixed at $10^{-3}$. Moreover, the wall-time is associated to the number of time steps, while, as seen below, the accuracy is mostly related with the integration time $\tau$, that is, the number of time steps multiplied by the discretization time interval. Fixing the discretization time interval makes them proportional. In our flow solver, the time step is upper bounded by the numerical stability of the solution, in the form of the Courant–Friedrichs–Lewy (CFL) condition, and the time integration truncation error at largest stable time step possible  is negligible in comparison with the space integration error.

\begin{table}[h!]
	\caption{Estimated coefficients of computational cost, in seconds}
	\label{tab:computationalCostCoeff}
	\centering
	\begin{tabular}{c|c|c}
		& Value (s) & Complexity \\
		\hline
		$C_\tau$ & $10$ & $\tau M$ \\
		$C_{It}$ & $0.6$ & $M$ \\
		$C_{GS}$ & $1.5\cdot10^{-4}$ & $M^2$ \\
		$C_{Eig}$ & $4\cdot10^{-10}$ & $M^3$ \\
	\end{tabular}
\end{table}

\begin{figure}[h!]
	\begin{center}
		\includegraphics[width=0.9\textwidth]{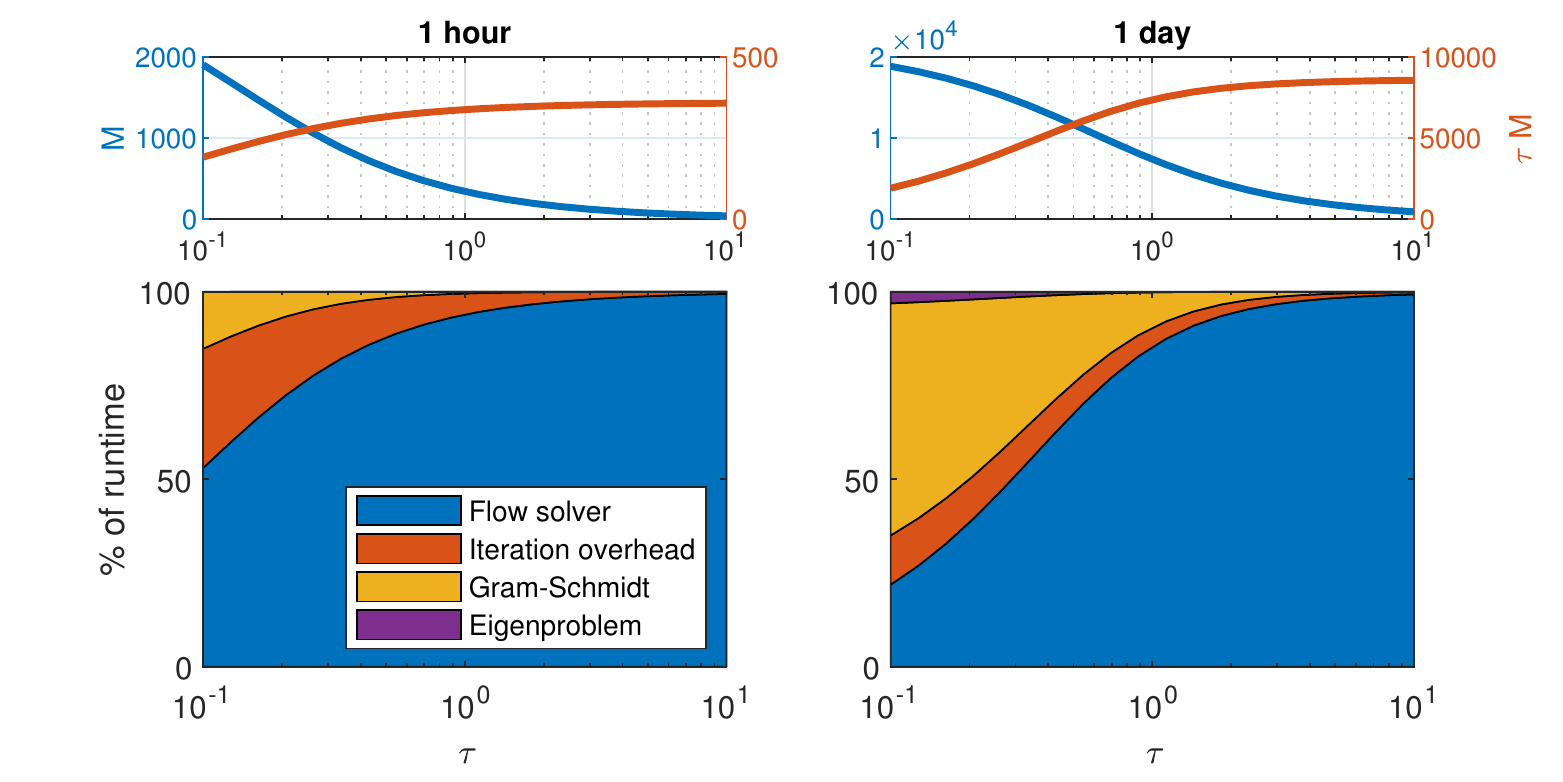}
	\end{center}
	\caption{(Top) Number of Arnoldi iterations ($M$) and total integration time ($\tau M$) as a function of integration time $\tau$ for two values of $C$. (Bottom) Fraction of time spent in each part of the procedure.}
	\label{fig:compCost}
\end{figure}

To illustrate these concepts, figure~\ref{fig:compCost}~(Top) shows the dimension $M$ of the problem that was possible to construct and solve for  $C$ = 1 hour and 1 day as a function of $\tau$. For a fixed $C$, $M$ increases as $\tau$ reduces. Also, as a function of $\tau$, the figure displays $\tau M$, here called total integration time, which increases with $\tau$. Figure~\ref{fig:compCost}~(Bottom) gives the fraction of $C$ spent on each part of the procedure. For $C$ = 1 hour or for larger values of $\tau$ with $C$ = 1 day, $M$ is small, and $C$ is dominated by the flow solver. As the Krylov space grows larger, the Gram-Schmidt normalization takes a greater portion of $C$, as it scales with $M^2$. Despite having a cost that scales with up to $M^3$, the time spent computing the eigenvalues of $H$ is still comparatively small in the parameter space searched. For smaller values of $\tau$ (thus larger $M$), the overhead of starting the flow solver also becomes more relevant. This plot considers a first-order accurate Fréchet derivative, but results are qualitatively similar for 2\textsuperscript{nd} and 4\textsuperscript{th}orders.

Restarted Arnoldi techniques \citep{Edwards1994,Lehoucq1996, Knoll2004, Bagheri2009}, allow the use of multiple smaller Krylov spans instead of a single larger one, which reduces the total memory usage of the algorithm as well as the cost components that scale super-linearly with $M$, such as the Gram-Schmidt orthonormalization. As will be seen, with optimal computational parameters these techniques tend to have a limited impact, and, hence, are not investigated here.

Another aspect that may be considered when analyzing the trade-off between $\tau$ and $M$ is memory usage. This algorithm requires storing all previous disturbances in memory in the $[\zeta_1 \dots \zeta_k]$ matrix, in our case up to around 2.4 MB per disturbance. However, memory optimization is of greater concern for the matrix-forming strategy \citep{Gennaro2013}, and is outside the scope of this paper.

\section{Flow solver and baseline results of the studied case}

Our test scenario is a flow over a two-dimensional open rectangular cavity. The Mach number is $Ma=0.5$, Reynolds number based on the cavity depth ($Re_D$) is 1000 and the boundary layer momentum thickness at the cavity leading edge is given by $D/\theta_0=100$. The cavity's aspect ratio is $L/D=2$. We chose this case as it is well documented and representative of typical cavity instability problems currently under study \citep{Bres2008, Yamouni2013, Mathias2021}. Under these conditions, this flow is unstable.

The flow solver used is described in more detail by Mathias and Medeiros\cite{Mathias2018}. It uses 4\textsuperscript{th} order compact spectral-like finite differences for spatial derivatives \citep{Bergamo2015} and a 4\textsuperscript{th} order Runge-Kutta for time integration. The code uses Cartesian grids and is parallelized by domain decomposition of the slab-pencil type \citep{Li2010,Martinez2015}. It solves the compressible Navier-Stokes equations. In this work the simulations are two-dimensional as this is enough for the paper purpose and limits the computational cost. Nevertheless, the conclusion is general and applicable to three-dimensional flows, as discussed in the last section.

\begin{figure}[h!]
	\begin{center}
		\includegraphics[width=0.7\textwidth]{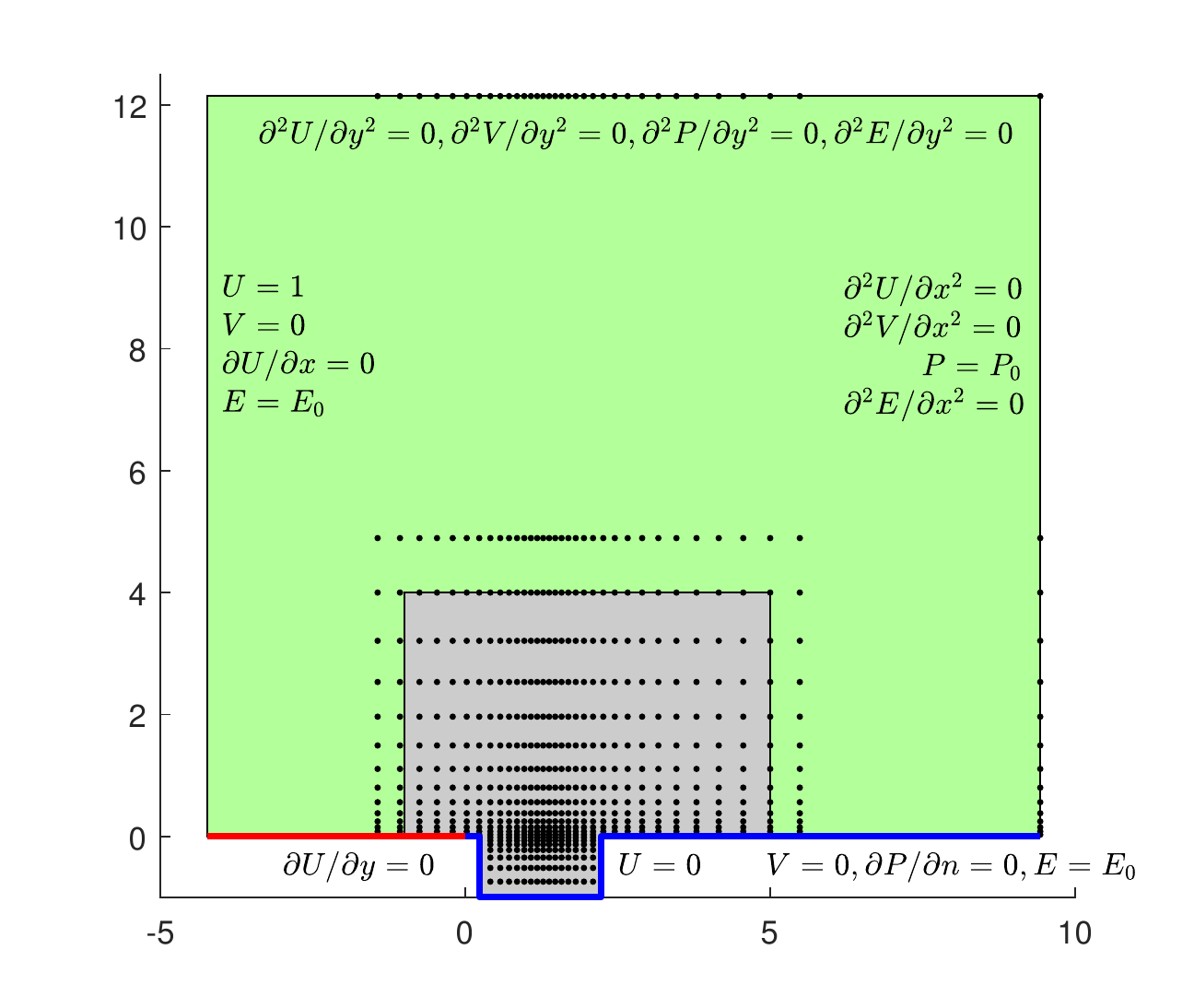}
	\end{center}
	\caption{Domain, mesh and boundary conditions of the flow simulation. The useful domain is marked in gray, while the buffer zone is green. One in every 10 nodes of the mesh is shown in black. The blue line represents wall boundaries, while the red line is a free-slip wall condition.}
	\label{fig:mesh}
\end{figure}

The mesh has $340 \times 220$ nodes in the stream-wise and wall-normal directions, respectively. The coordinate system origin is at the beginning of the boundary layer, while the cavity leading edge is at (0.23,0). All lengths are normalized by the cavity depth and the density and velocities, by the free-stream values. The useful domain spans from $x=-1$ to $5$ in the stream-wise direction and from $y=-1$ to $4$ in the wall-normal direction. At the open boundaries of the domain, there is a buffer zone, which progressively increases node spacing and lowers differentiation accuracy order, designed to prevent boundary effects from contaminating the solution. Upstream of the origin, a free-slip condition on the wall was used, which better negotiates the stagnation point at the origin \citep{Martinez2016}. Figure~\ref{fig:mesh} shows the domain, the mesh and the boundary conditions. Both base flow and stability results were independent on the computational mesh and domain.

\begin{figure}[h!]
	\begin{center}
		\includegraphics[width=0.7\textwidth]{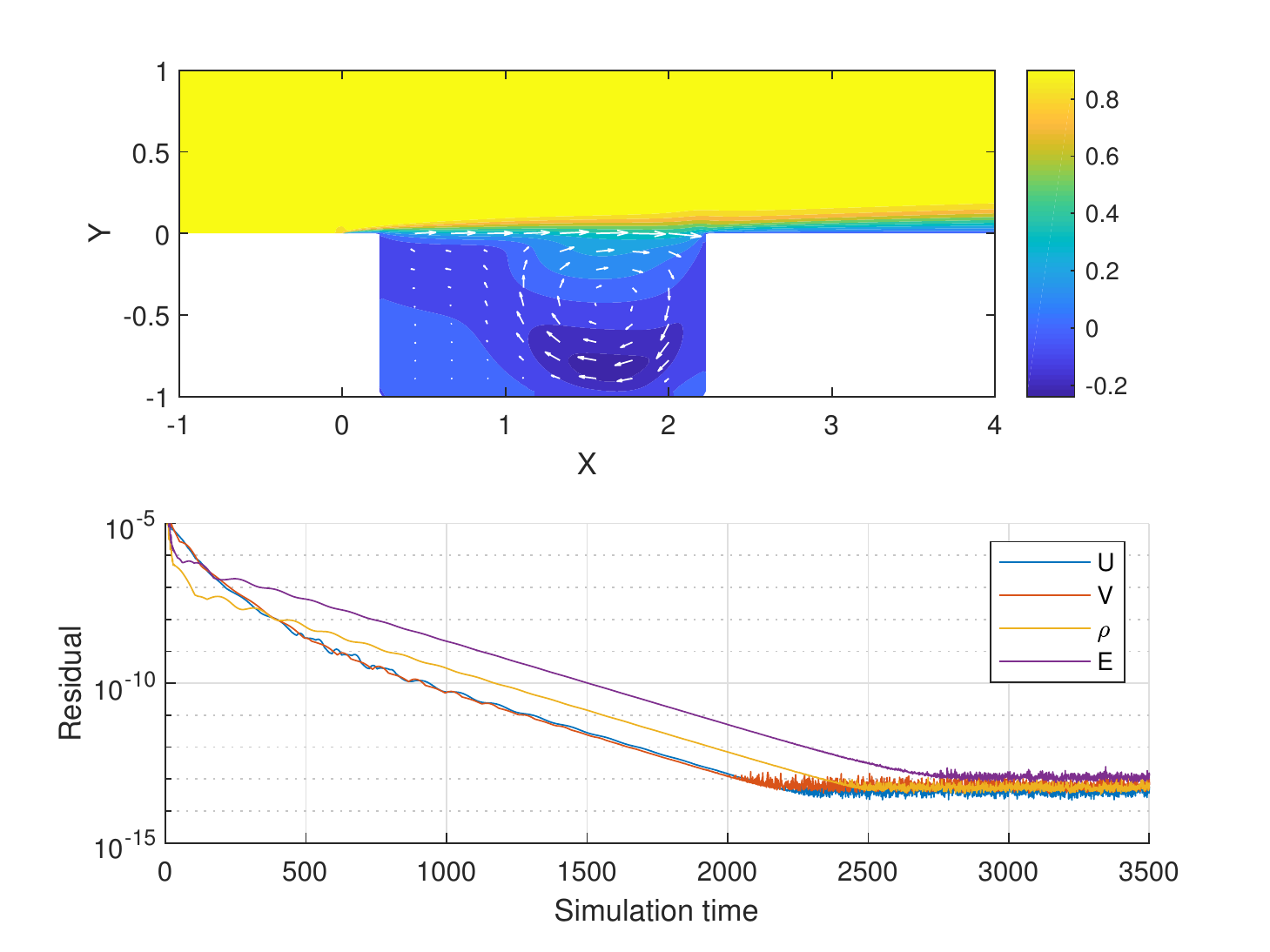}
	\end{center}
	\caption{(Top) Streamwise velocity of the base flow (contours every 10\% of the free-flow, from -30\% to 90\%) and velocity field inside the cavity. (Bottom) Maximum residue of each variable as the flow evolves to the steady state.}
	\label{fig:baseflow}
\end{figure}

The base flow was obtained using the flow solver with the aid of a Selective Frequency Damping (SFD) filter, which damps modes below a certain frequency threshold \citep{Akervik2006}. Without such filter, the unstable modes of the flow would prevent reaching a steady state. The SFD filter was turned off during the stability analysis. Figure~\ref{fig:baseflow}~(Top) depicts the base flow in the cavity region, while figure~\ref{fig:baseflow}~(Bottom) shows the base flow convergence history. All residuals are in the order of $10^{-13}$ relative to free-flow values.

\begin{figure}[h!]
	\begin{center}
		\includegraphics[width=0.9\textwidth]{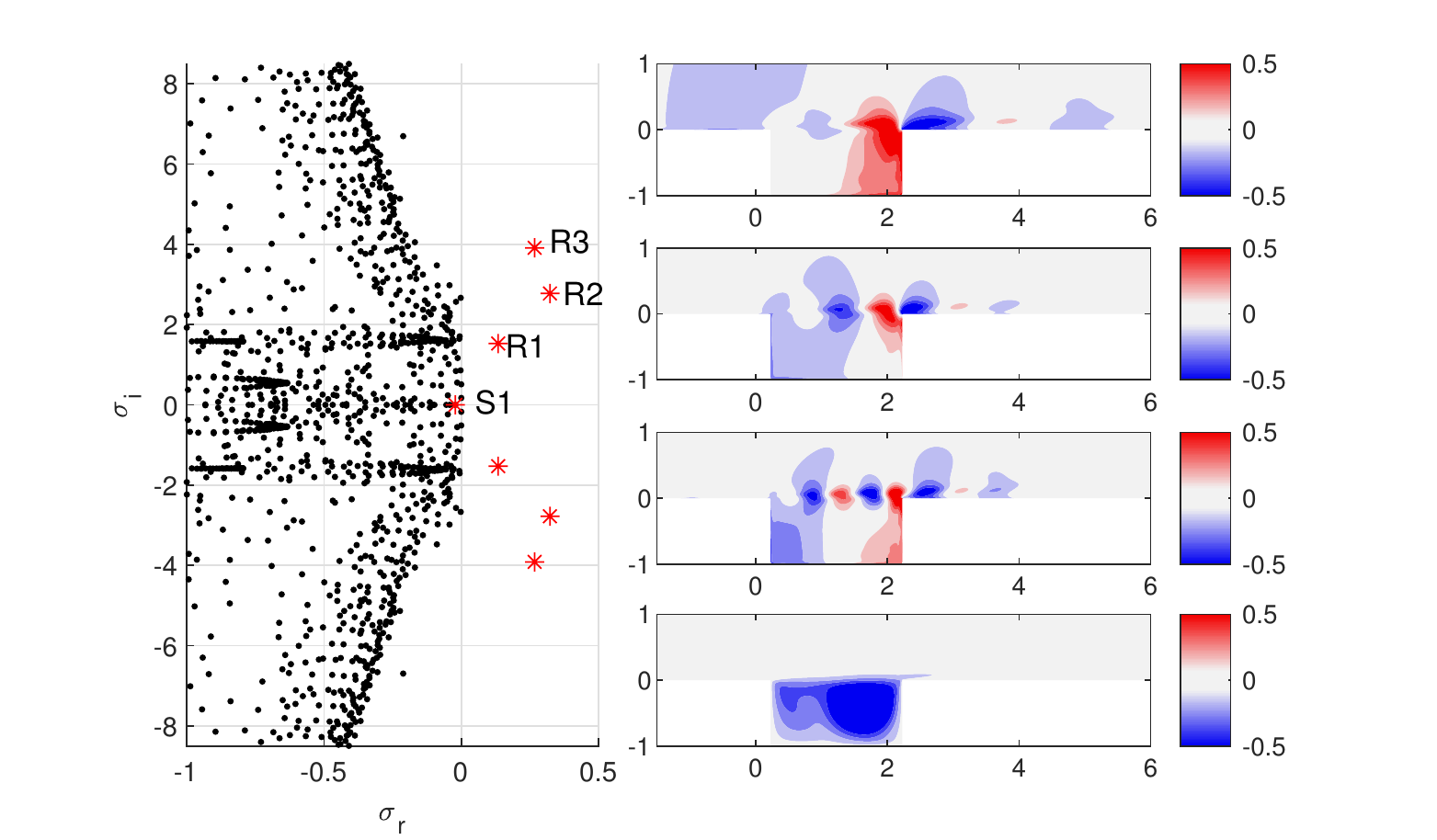}
	\end{center}
	\caption{(Left) Eigenspectrum of the base flow, with the unstable Rossiter modes (Rn) and the least stable stationary mode (S1) indicated, (Right) Density contours for the Rossiter modes 1 to 3 and the Stationary mode 1.}
	\label{fig:modesExample}
\end{figure}

Using the techniques described, the global stability analysis of this flow was performed. Figure~\ref{fig:modesExample}~(Left) shows the eigenvalue spectrum. A positive real part indicates an unstable mode. There are six unstable modes, known as Rossiter modes ($\sigma_{R1}=0.1344\pm1.5276i$, $\sigma_{R2}=0.3230\pm2.7762i$ and $\sigma_{R3}=0.2672\pm3.9143i$) \citep{Rossiter1964, Mathias2021}, and their respective complex conjugates. Also indicated is mode S1 ($\sigma_{S1}=-0.0215\pm0i$) the least stable stationary mode, i.e., with a null imaginary part. It relates to the transport of internal energy in the cavity \citep{Bergamo2015}. The right-hand side of the figure shows density contours of these modes, the Rossiter modes at an arbitrary phase. In the following sections, we will use as examples mode R2, the most unstable for this set of parameters and mode S1, which is slightly stable. Their stability results will illustrate the sensitivity analysis and optimal value estimates presented in the following sections.

\section{Order of accuracy of the Fréchet derivative and disturbance magnitude}
\label{sec:epsilon}

The Fréchet derivative approximates the product $B \zeta$ for the Arnoldi algorithm. In this section we focus on two choices that must be made: the truncation order of the derivative approximation and the magnitude of the disturbance, $\varepsilon$. On the one hand, reducing the disturbance magnitude reduces the truncation error in the Fréchet derivative. On the other hand, owing to the discretization error in the flow solver, reducing the disturbance may deteriorate the approximations of $F$, as it involves the evolution of disturbances that are several orders of magnitude smaller than the base flow. Moreover, stable modes may attenuate to orders of magnitude smaller than their original amplitude. These aspects advocate the use of high accuracy order methods \citep{Lele1992, Gaitonde1998, Souza2005, Silva2010}. Often, the flow equations include the nonlinear terms. This may pose another upper bound for $\varepsilon$, but this is strongly dependent on $\tau$ and hence is postponed to the next section.

If the truncation error of the Fréchet derivative becomes a limiting factor for the disturbance magnitude, a higher order approximation must be used. This brings the extra computational cost of increasing the number of calls to the flow solver. As an example, Tezuka and Suzuki\cite{Tezuka2006} used the second-order method and noted that the effect of changing $\varepsilon$ within two orders of magnitude was negligible. This indicates that the discretization error of the flow solver, the nonlinear terms, and the truncation error of the Fréchet derivative were negligible within the range of parameters of that study.

In our study, we tested disturbance magnitudes in the range $10^{-11}~\le~\varepsilon~\le~10^{-3}$, combined with different values of $\tau$. Both first and second-order accurate Fréchet derivatives were evaluated. The Krylov span $M$ was so large that, within machine accuracy, the results for the eigenvalues tracked did not change for larger spans.

\begin{figure}[h!]
	\begin{center}
		\includegraphics[width=0.9\textwidth]{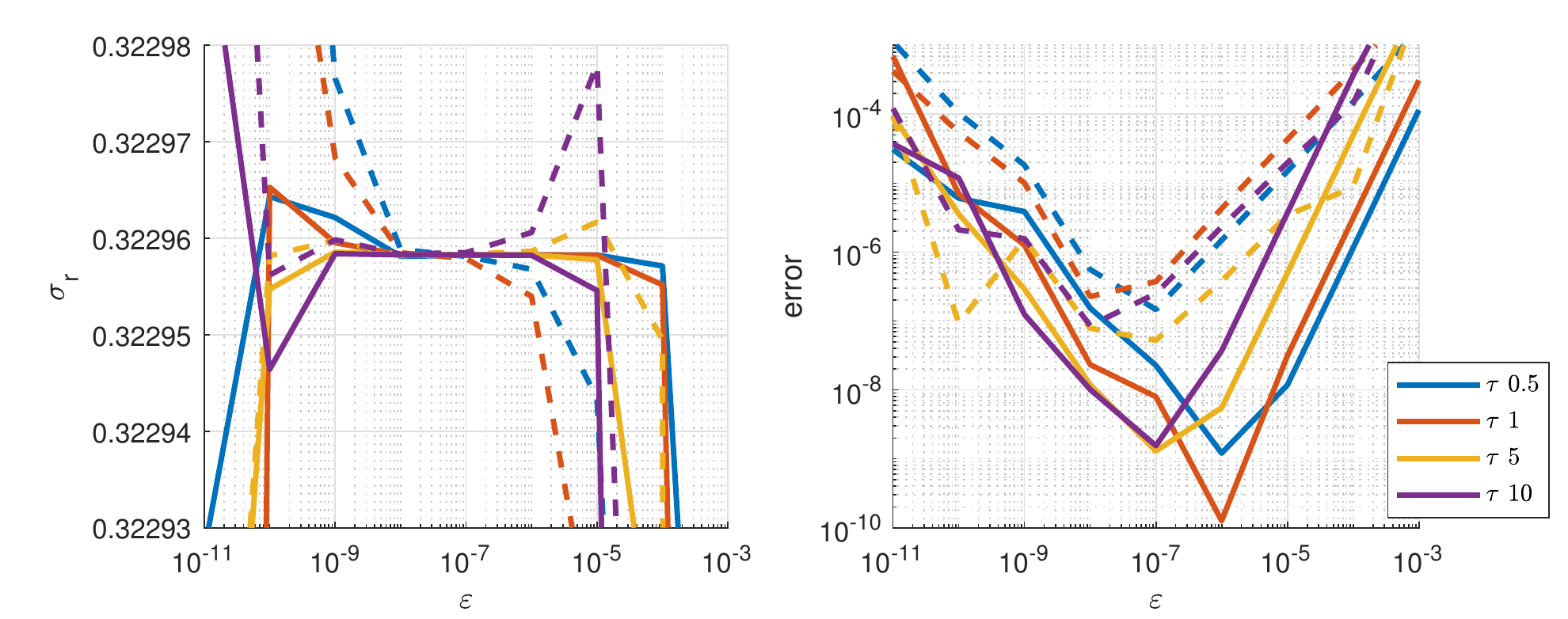}
	\end{center}
	\caption{Real part of the R2 eigenvalue as a function of $\varepsilon$ for different values of $\tau$. Dashed lines represent first-order accurate Fréchet derivative, and full lines, second-order accurate. (Left) Absolute value. (Right) Difference to the reference case.}
	\label{fig:epsConvMode1}
\end{figure}

Figure~\ref{fig:epsConvMode1}~(Left) shows the values of the real part of the R2 eigenvalue obtained with different combinations of parameters. The results are in better agreement with each other in the range $10^{-9} \le \varepsilon \le 10^{-6}$. For reference, we performed calculations with $\tau=1$, $\varepsilon = 10^{-6}$, $M=2500$ and a fourth-order accurate Fréchet derivative. This Fréchet derivative is more expensive, but provides more accurate results. Throughout this study, this case will be called reference case. Figure~\ref{fig:epsConvMode1}~(Right) shows the difference between each case and this reference.

From figure~\ref{fig:epsConvMode1}~(Right), in comparison with first-order accurate Fréchet, the second-order one considerably improved the accuracy. For large $\varepsilon$ the truncation error dominates and the error decreases as $\varepsilon$ diminishes. Accordingly, this is the region most affected by the Fréchet derivative truncation order. Close inspection reveals that, in this region, the error for $\sigma_r$ decays approximately with either first or second-order, depending on the Fréchet derivative approximation used. This lends support to the approach here used and demonstrates the reference case represents an effective exact solution for the problem.

Also from figure~\ref{fig:epsConvMode1}~(Right), for small $\varepsilon$, the error grows as $\varepsilon$ decreases. This is associated with inaccuracy in the $F$ approximations and originates in the flow solver discretization error. Accordingly, in this region the error is less affected by the Fréchet derivative truncation order. There, the error scales roughly linearly with $1/\varepsilon$, which is the expected scale for this type of error in the estimates of a first derivative like the Fréchet one. For first-order accurate Fréchet derivative the error scaling produces the characteristic "V" shape of the error curve in figure~\ref{fig:epsConvMode1}~(Right).

The maximum accuracy is better than $10^{-8}$, but the optimum for $\tau=1$ is likely to be fortuitous owing to the greater similarity of parameters with the reference case. The minimum error for 2\textsuperscript{nd} order accuracy is about two orders of magnitude lower than for first-order accuracy; however, if an accuracy of $10^{-6}$ were acceptable, first-order would have produced sufficiently accurate results at about half the cost. For first-order accurate, the optimum $\varepsilon$ is between $10^{-8}$ and $10^{-7}$. Consistent with the error scaling discussed above, the optimum $\varepsilon$ for second-order accuracy is larger than that of first-order accuracy. The optimum value of $\varepsilon$ is lower for larger values of $\tau$ due to the nonlinear effects to be discussed in section~\ref{sec:tau}.

\begin{figure}[h!]
	\begin{center}
		\includegraphics[width=0.9\textwidth]{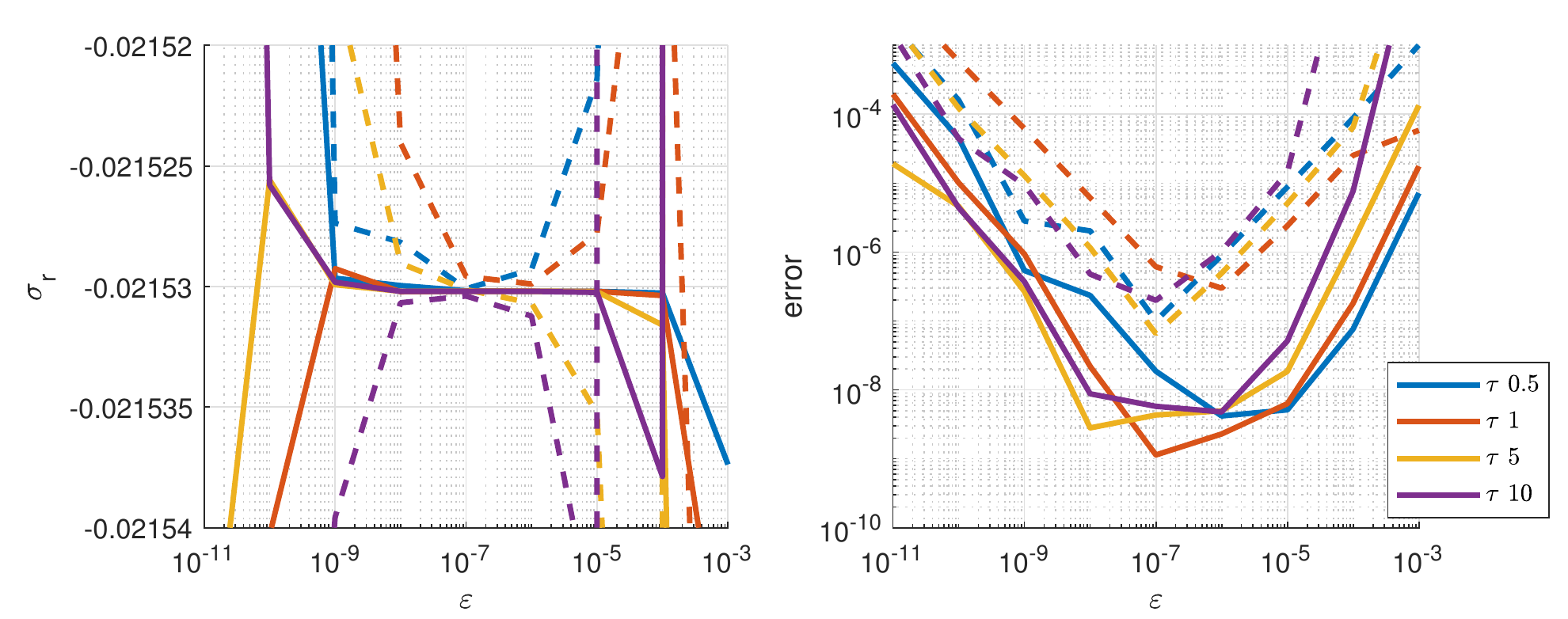}
	\end{center}
	\caption{Real part of the S1 eigenvalue  as a function of $\varepsilon$ for different values of $\tau$. Dashed lines represent first-order accurate Fréchet derivative, and full lines, second-order accurate. (Left) Absolute value. (Right) Difference to the reference case.}
	\label{fig:epsConvModeT}
\end{figure}

Figure~\ref{fig:epsConvModeT} shows the result for mode S1. In general, the same conclusions are reached. Being a stable mode, S1 is more vulnerable to discretization errors in the flow solver. Hence the effect of the discretization error is felt earlier for this mode, slightly raising the minimum error.

Despite the complexity, the error behavior closely follows the theoretical error scaling in the asymptotic range of discretization \citep{Silva2010}. It is possible to predict the order of magnitude of the optimal $\varepsilon$ for a first order accurate Fréchet derivative approximation by replacing the exact solution $F$ by the numerical solution of the solver in equation~\ref{eq:taylor3}. This introduces all the errors intrinsic to the flow solver. Using $\tilde{F}$ to indicate the numerical solution,

\begin{equation}
	\tilde{F}(U_0 + \varepsilon_0 \zeta) = \tilde{F}(U_0) + \varepsilon_0 B \zeta + \overrightarrow{\mathcal{O}}\left(\Vert \varepsilon_0^2 \zeta \Vert\right)\ + \overrightarrow{E_F} \ ,
	\label{eq:error0}
\end{equation}

\noindent where $\overrightarrow{E_F}$ is the error of the numerical approximation of $F$. Isolating $B \zeta$ (remember that $\norm{\zeta}=1$),

\begin{equation}
	{B} {\zeta} = \frac{1}{\varepsilon_0} \left[ {\tilde{F}}\left({U}_0 + \varepsilon_0 {\zeta}\right) - {\tilde{F}}\left({U}_0\right) \right] + \overrightarrow{\mathcal{O}}(\varepsilon_0) + \frac{\overrightarrow{E_F}}{\varepsilon_0} \ .
	\label{eq:frechet1e}
\end{equation}

\noindent It is possible to derive similar expressions for Fréchet derivatives of arbitrary accuracy order $n$, which we can write in general form as:

\begin{equation}
	{B} {\zeta} = \delta_n \tilde{F} + \overrightarrow{\mathcal{O}}(\varepsilon_0^n) + \frac{\overrightarrow{E_F}}{\varepsilon_0} \ ,
	\label{eq:frechet1f}
\end{equation}

\noindent where $\delta_n \tilde{F}$ is the $n$\textsuperscript{th} order accurate finite differences approximation for the Fréchet derivative.

The numerical error of $F$, $\overrightarrow{E_F}$, has two main components: the discretization errors of the flow solver, and the errors caused by nonlinear effects in the flow solution. However, as anticipated, the nonlinear error is neglected in the analysis here and considered in the next section. Each scalar component in the vector $\overrightarrow{E_F}$ has a discretization error of the order of $\varepsilon_S$, which is a measure of the flow solver's error. Therefore, the norm of $\overrightarrow{E_F}$ is $\varepsilon_S \sqrt{N}$, where $N$ is the total number of variables in the system. Replacing $\varepsilon_0$ by $\varepsilon \sqrt{N}$, the total error of $B \zeta$ in equation~\ref{eq:frechet1f} is estimated as

\begin{equation}
	E_B = \left\Vert \overrightarrow{\mathcal{O}}(\varepsilon_0^n) + \frac{\overrightarrow{E_F}}{\varepsilon_0} \right\Vert = \mathcal{O}\left( (\varepsilon \sqrt{N})^n\right) + \mathcal{O}\left(\varepsilon_S/\varepsilon\right) \ .
	\label{eq:error3}
\end{equation}

\noindent The optimum $E_B$ is where $\partial E_B/ \partial \varepsilon$ is null, leading to

\begin{equation}
	\varepsilon_{opt} = \mathcal{O} \left( \left( \frac{\varepsilon_S}{ N^{n/2}}  \right) ^ {\frac{1}{n+1}} \right)\ .
	\label{eq:error4}
\end{equation}

With $\varepsilon_S=10^{-13}$, from figure~\ref{fig:baseflow}, and $N=245340$, we can estimate the minimum error for first-order accurate Fréchet derivative as $\varepsilon_{opt}=\mathcal{O}(10^{-8})$ and for the second-order accuracy as $\varepsilon_{opt}=\mathcal{O}(10^{-6})$. The error scaling involves coefficients which were disregarded, yet the estimates are close to what was observed in figures~\ref{fig:epsConvMode1}~and~\ref{fig:epsConvModeT}.

It is important to note that the theoretical analysis estimates the error in $B \zeta$ while figures~\ref{fig:epsConvMode1}~and~\ref{fig:epsConvModeT} show the error for selected eigenvalues. Moreover, this equation does not account for non-linear errors, which  occur if both $\varepsilon$ and $\tau$ are large enough. These issues are best discussed in connection with $\tau$ and are, hence, left to the next section. In any case, as will be seen, following the guidelines for optimal parameters, the nonlinear error becomes negligible and equation~\ref{eq:error4} is enough to determine the optimal value of $\varepsilon$.

\begin{figure}[h!]
	\begin{center}
		\includegraphics[width=\textwidth]{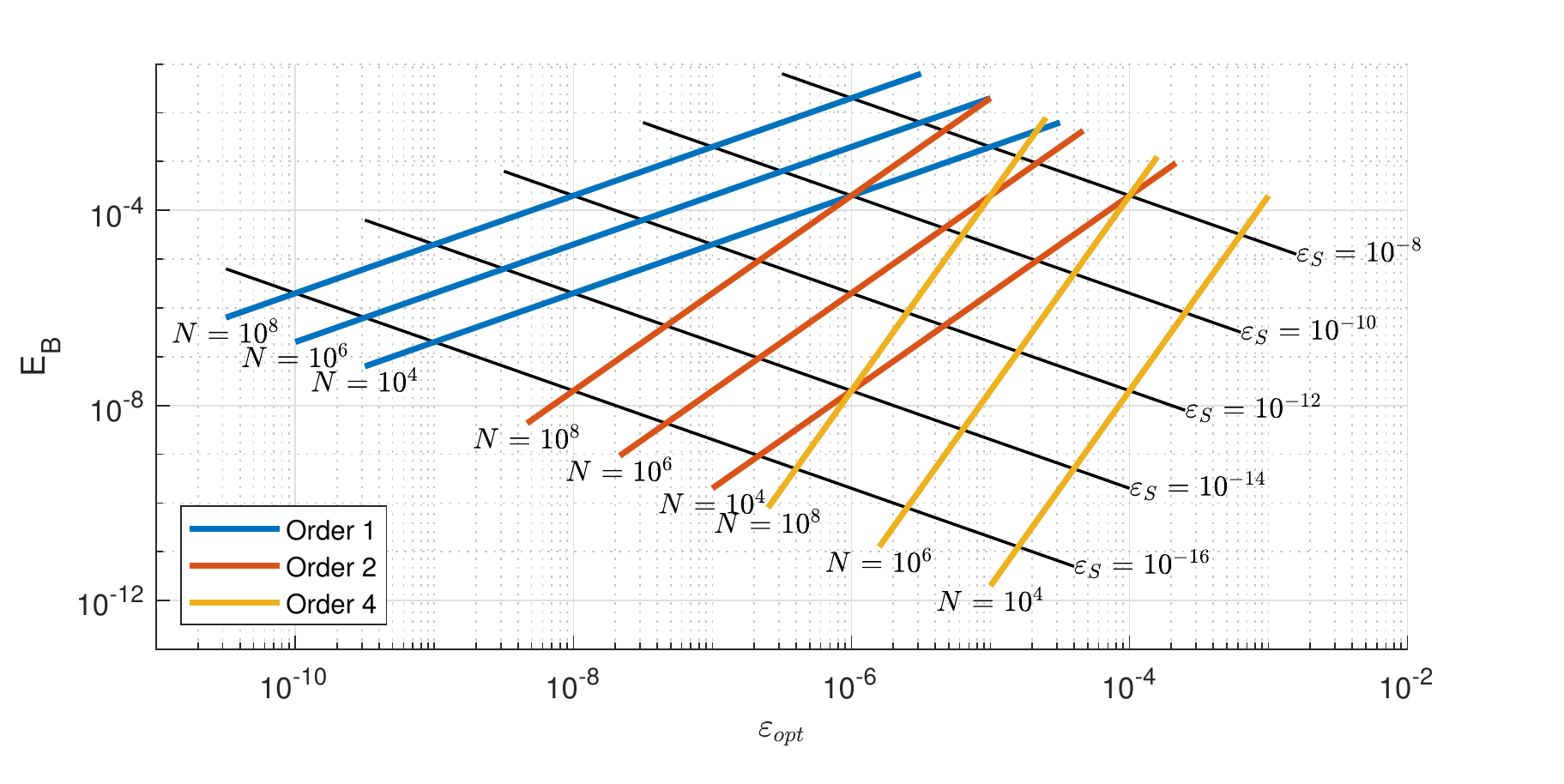}
	\end{center}
	\caption{Values of $\varepsilon_{opt}$ and the corresponding $E_B$ as a function of $\varepsilon_S$ and $N$ for 1\textsuperscript{st}, 2\textsuperscript{nd} and 4\textsuperscript{th} order accuracy Frèchet approximations.}
	\label{fig:epsOpt}
\end{figure}

According to equations~\ref{eq:error3}~and~\ref{eq:error4}, the size of the grid, the discretization error of the flow solver, and the truncation order of the Fréchet derivative approximation affect the optimal disturbance amplitude and the minimal error $E_B$ in different ways. These dependencies are explored in figure~\ref{fig:epsOpt}. Regarding the error $E_B$, higher order Fréchet derivative approximations provide better results, which deteriorate faster with an increase in $N$ or $\varepsilon_S$. For example, if $\varepsilon_S=10^{-16}$ and $N=10^4$, going from first order accurate to fourth order accurate Fréchet derivative reduces $E_B$ more than $10^4$ times. For a less accurate code and larger grids, say $\varepsilon_S=10^{-8}$ and $N=10^8$, the same change reduces $E_B$ by only a factor of $10$, even at optimal conditions. This warns that numerical errors associated with less accurate codes in large grids cannot be compensated by higher order accurate Fréchet derivative schemes. About $\varepsilon_{opt}$, high-order accurate Fréchet derivative approximations require a substantially larger $\varepsilon$ for optimal performance. Moreover, for low-order accurate Fréchet derivatives, $\varepsilon_S$ has a much larger impact on $\varepsilon_{opt}$. This indicates that, for a given grid, the use of high order accurate Fréchet derivatives requires a finer tuning to achieve optimal conditions. On the other hand,  for low order accurate Fréchet derivative approximation, the optimal conditions are less sensitive to the grid size $N$.

\section{Effect of integration time length}
\label{sec:tau}

Another parameter to be chosen when solving the Fréchet derivative to approximate $B \zeta$ is the integration time $\tau$. A larger $\tau$  can better isolate one mode from the others within the Arnoldi algorithm; on the other hand, if the integration time is too long, unstable modes may grow sufficiently to develop nonlinear effects, and stable modes may be attenuated enough as to be contaminated by the discretization error of the solver. Small $\tau$ also reduces the accuracy of the Fréchet derivative approximations. In this section, we compute the overall error for various values of $\tau$ and develop equations for optimal $\tau$. According to Goldhirsch et al.\cite{Goldhirsch1987}, every time the flow solver is called, it acts in the direction of filtering the most unstable (or least stable) modes present in the flow; they introduced the concept of filtering time, $\tau M$, which we refer to as total integration time. They defined a lower limit for $\tau M$ as

\begin{equation}
	\tau M \gg \frac{1}{\abs{ \sigma_i^R - \sigma_{M+1}^R} } \ ,
	\label{eq:lowerBound}
\end{equation}

\noindent where $\sigma_{M+1}$ is the eigenvalue to be retrieved in the next iteration and $\sigma_i^R$ is the real part of the sought eigenvalue, $M+1>i$. For a fixed $M$, equation \ref{eq:lowerBound} is a lower bound for $\tau$. According to Gómez et al.\cite{Gomez2014}, there is also an upper bound of $\tau$ for a stable mode as:

\begin{equation}
	\tau \ll \frac{\ln{\left(\frac{\varepsilon_{S}}{\varepsilon}\right)}}{\sigma_i^R} \ ,
	\label{eq:upperBound1}
\end{equation}

\noindent where $\varepsilon$ is the magnitude of the disturbance added to the base flow and $\varepsilon_{S}$ is the flow solver's discretization error. This limit is associated with the amplitude of the stable mode becoming too small and hence vulnerable to the flow solver’s discretization errors.

\begin{figure}[h!]
	\begin{center}
		\includegraphics[width=0.8\textwidth]{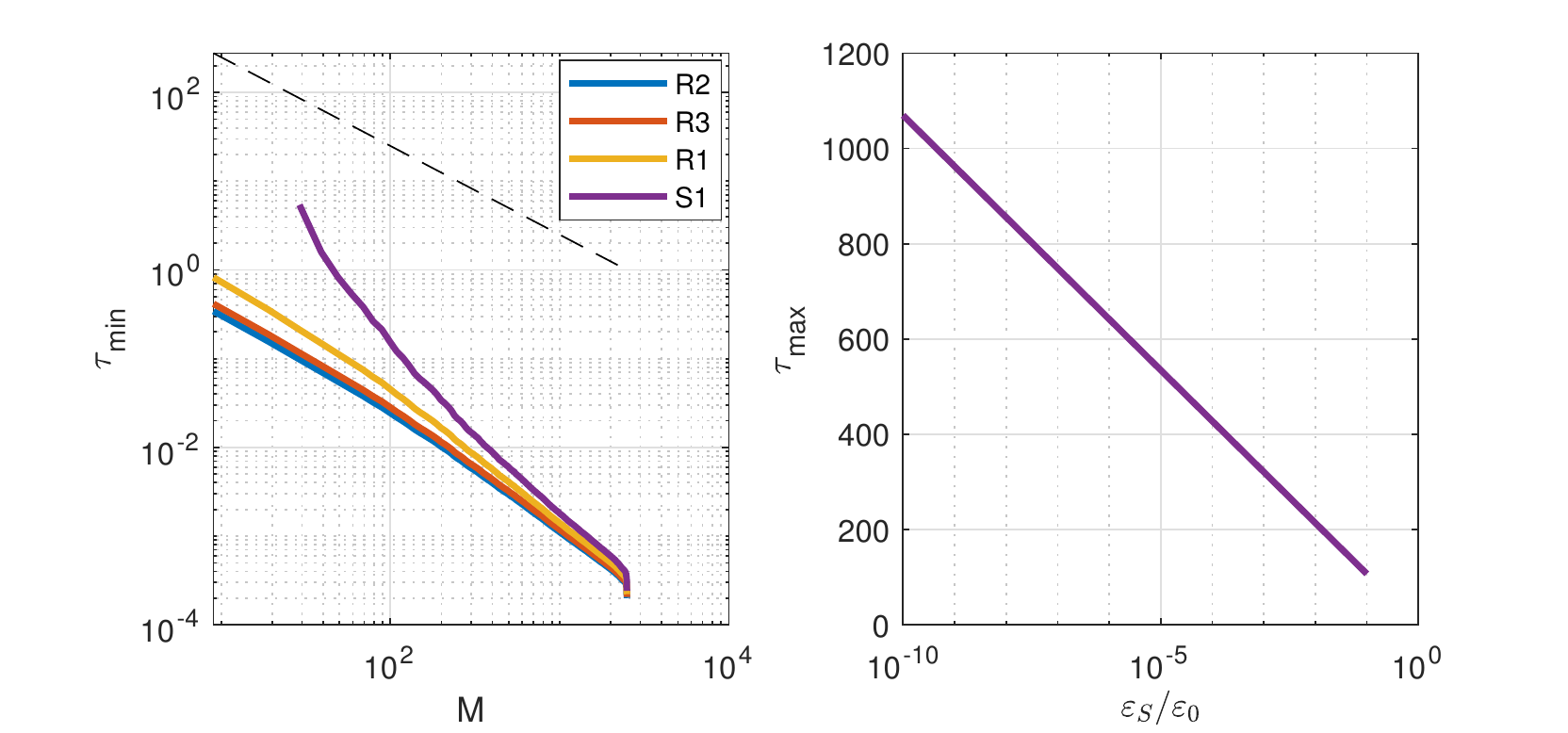}
	\end{center}
	\caption{(Left) Lower limit for $\tau$ according to equation~\ref{eq:lowerBound}. The dashed line represents the product $\tau M = 2500$, which was used in this work. (Right) Upper limit for $\tau$ according to equation~\ref{eq:upperBound1}.}
	\label{fig:tauBounds}
\end{figure}

Figure~\ref{fig:tauBounds} gives both (Left) lower and (Right) upper bounds for $\tau$ according to equations~\ref{eq:lowerBound}~and~\ref{eq:upperBound1}. Lower limits for modes R1, R2, R3 and S1 are presented, but only S1 has its upper limit shown, as equation~\ref{eq:upperBound1} is only valid for stable modes. In our tests, $\tau M=2500$ (dashed line), which places our simulations well above the lower bounds for all modes considered. The upper bound equation requires $\varepsilon_S$, which is estimated as $10^{-13}$, from figure~\ref{fig:baseflow}. The maximum $\tau$ used in our study was 50, which places our simulations well below this bound for the range $10^{-11}<\varepsilon<10^{-3}$ tested.

In practice, however, more restrictive bounds often apply. Moreover, the bounds do not indicate optimum values. To investigate this, we used results similar to those of the previous session, but presented in a way to better explore the sensitivity of the performance to the integration time.

As seen above, $\tau$ can affect the Krylov span approximation as well as the $F$ and the Fréchet derivative approximations. In this section we address only the effect of $\tau$ on the $F$ and the Fréchet derivative approximations and therefore we ensured that $M$ was so large as not to affect the results. The effects of $M$ and $\tau$ on the Krylov span approximation are discussed in the next section.

\begin{figure}[h!]
	\begin{center}
		\includegraphics[width=0.9\textwidth]{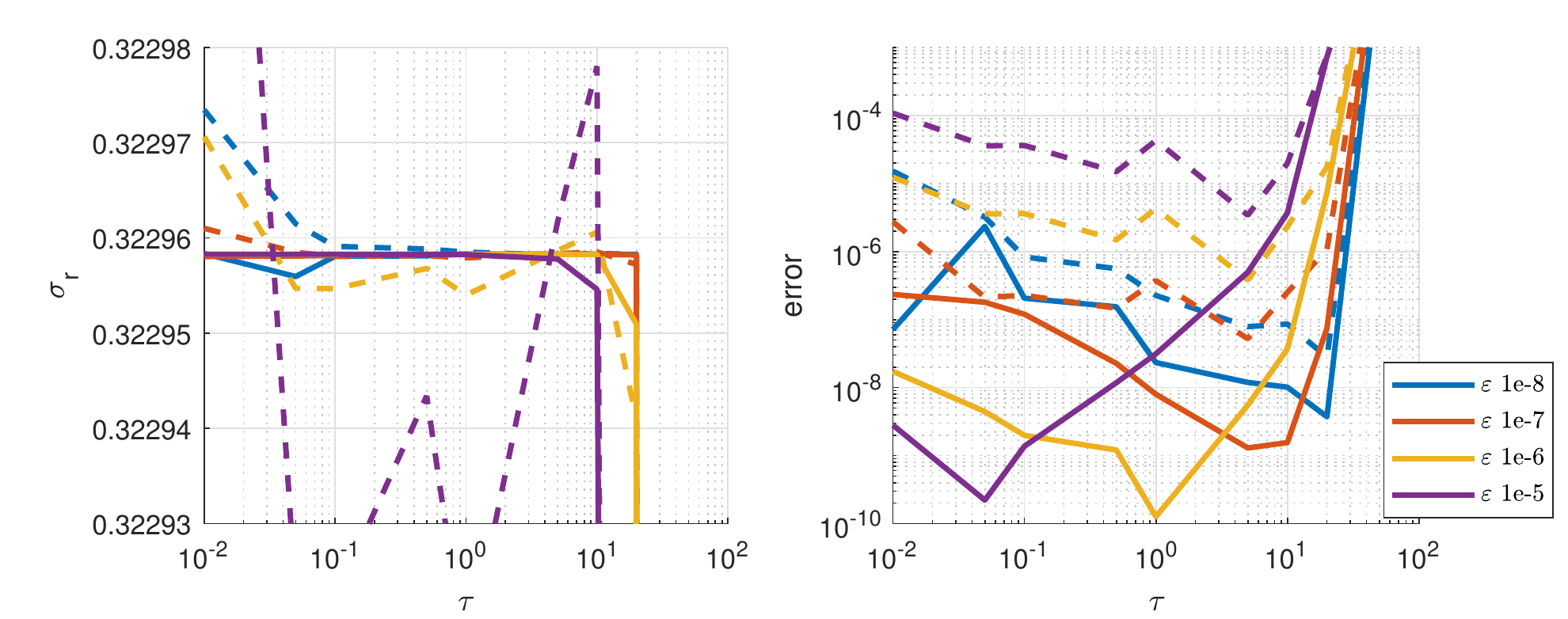}
	\end{center}
	\caption{Real part of the R2 eigenvalue as a function of $\tau$ for selected values of $\varepsilon$. Dashed lines represent the first-order accurate Fréchet derivative, while full lines represent the second-order accurate derivative. (Left) Absolute value. (Right) Difference to the reference case.}
	\label{fig:dtConvMode1}
\end{figure}

\begin{figure}[h!]
	\begin{center}
		\includegraphics[width=0.9\textwidth]{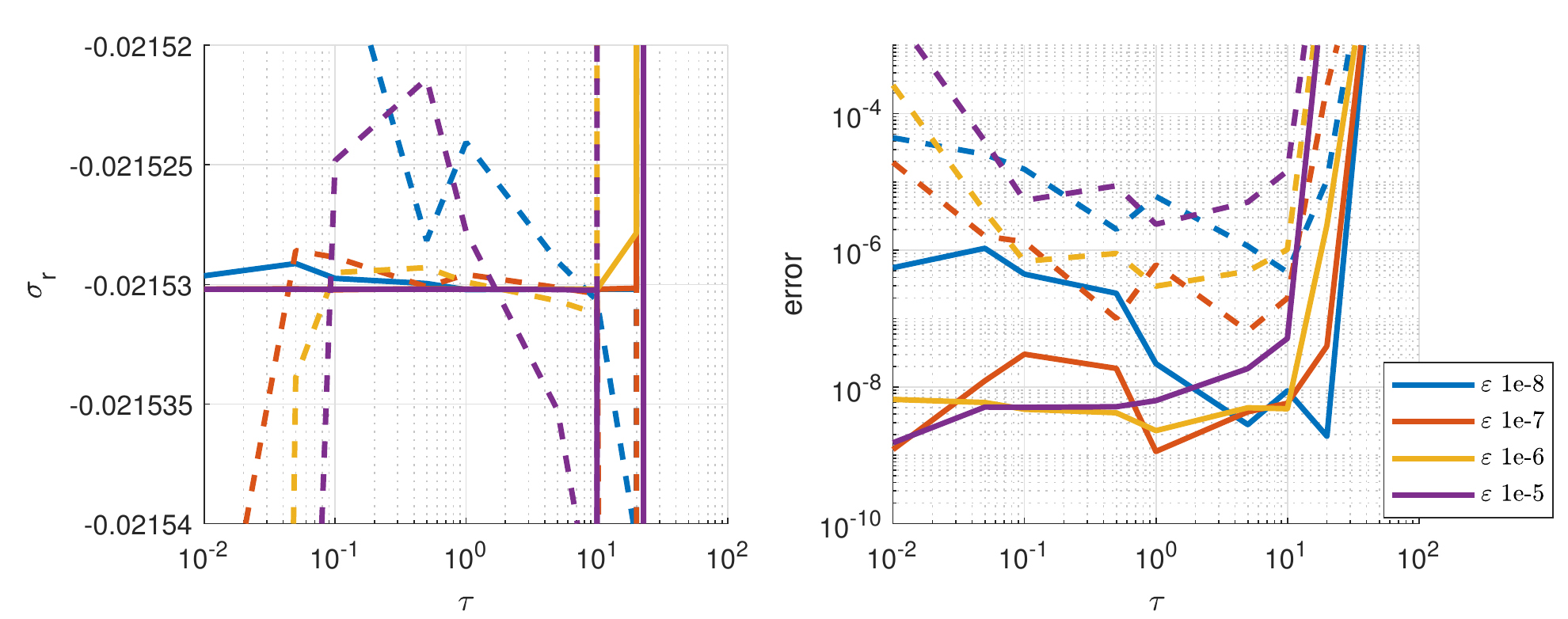}
	\end{center}
	\caption{Real part of the S1 eigenvalue as a function of $\tau$ for selected values of $\varepsilon$. Dashed lines represent the first-order accurate Fréchet derivative, while full lines represent the second-order accurate derivative. (Left) Absolute value. (Right) Difference to the reference case.}
	\label{fig:dtConvModeT}
\end{figure}

Figure~\ref{fig:dtConvMode1} shows the real part of the R2 eigenvalue for various combinations of $\tau$ and $\varepsilon$, with both first and second-order accurate Fréchet derivatives. It also shows the error relative to the reference case. Figure~\ref{fig:dtConvModeT} shows the same data, but for mode S1. A rapid increase in error is observed on the right-hand side of these figures. This is associated with nonlinear effects in the flow solution as we will discuss. Note, the effective upper boundaries of $\tau$ are much more restrictive than those of equation~\ref{eq:upperBound1} even for the stable modes. The error increase for lower $\tau$ cannot be linked to the effect described by equation~\ref{eq:lowerBound}, as we have chosen a value of $M$ large enough so that it no longer influences the results. This error has a different origin which will also be addressed.

\begin{figure}[h!]
	\begin{center}
		\includegraphics[width=0.8\textwidth]{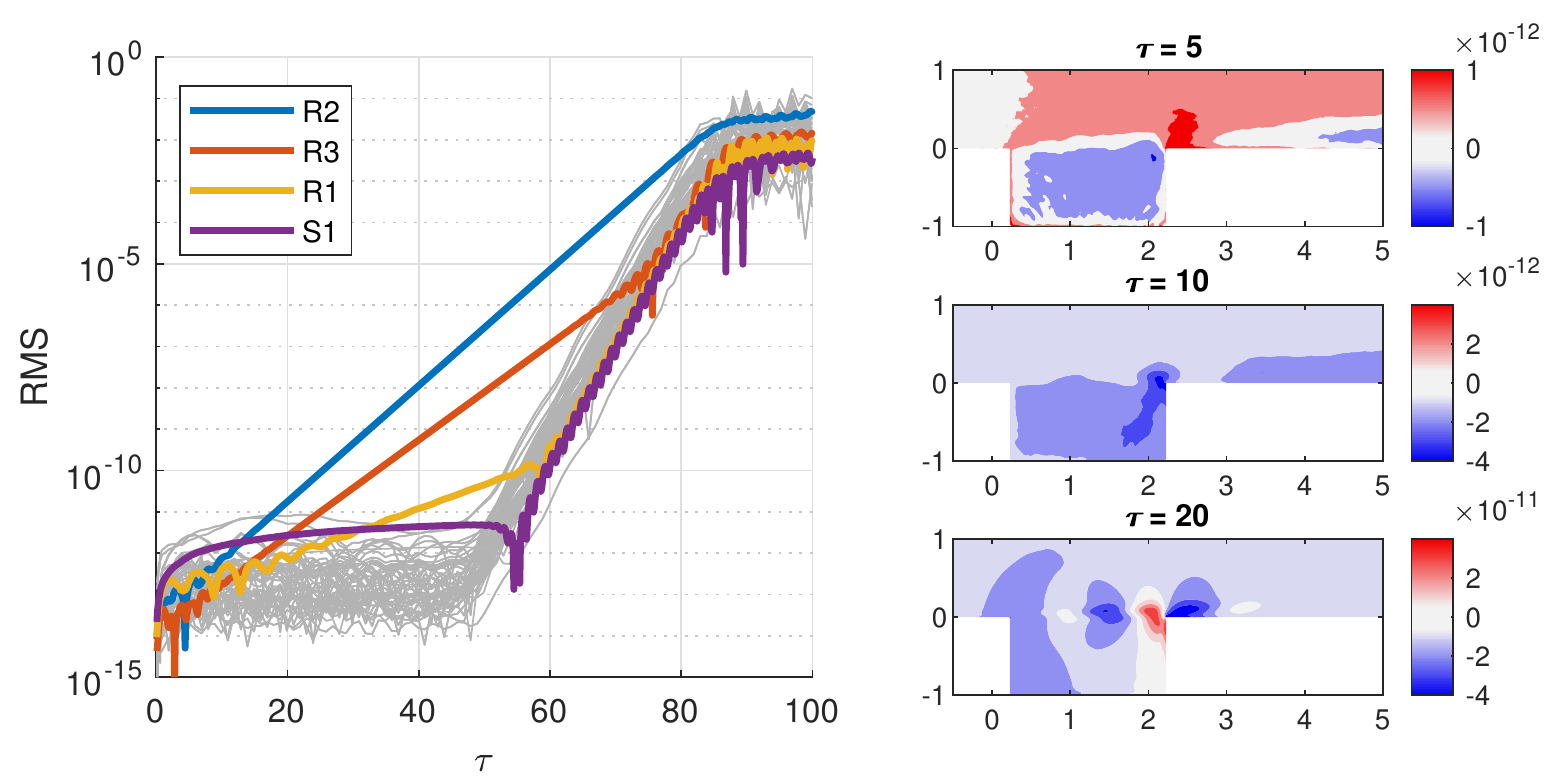}
	\end{center}
	\caption{(Left) Evolution of each mode in the flow as it is simulated starting from the base flow without the introduction of any disturbance other than the discretization error. Modes R1 to R3 and S1 are highlighted. Solution projections on the other 100\textsuperscript{th} least stable are shown as thin gray lines. (Right) Snapshots of the density fluctuation at $\tau=5$, $10$, and $20$.}
	\label{fig:modesInBaseflow}
\end{figure}

To explain the nonlinear effect, we use figure~\ref{fig:modesInBaseflow}. On the left, the figure shows results of the time evolution of the base flow with no disturbances other than the noise floor, which is at the level $10^{-13}$. The figure displays the contribution of each mode to the flow as it evolves. These curves were obtained by reconstructing the output from the flow solver from linear combination of all 2500 modes retrieved by the global stability analysis. It is impossible to fully reconstruct the signal unless all modes are available, so we have used a least squares approach to obtain an optimal representation in this limited set. At $\tau=20$, mode R2 already dominates the flow. However, as early as $\tau=10$, modes R2 and R3 are present and growing exponentially. Figure~\ref{fig:modesInBaseflow}~(Right) shows snapshots of the flow fluctuations along the time evolution, illustrating the emergence of the R2 mode. At $\tau=50$, the gray lines that depict stable modes start growing as well, until reaching saturation at $\tau=90$; yet, those modes did not become unstable. This is an artifact of the dominant mode's nonlinearity.

An estimate of the magnitude of the nonlinear terms of the disturbance Navier-Stokes equations is the square of the leading mode, formally $\varepsilon^2 e^{ 2 \sigma_1^R \tau}$, which matches the growth rate observed in the figure. As this nonlinear content grows, it contaminates the evolution of the sought modes because the Arnoldi algorithm projects this nonlinear content onto the linear modes. Even the leading mode R2 is affected, as reflected in the irregular oscillations shortly after the limit cycle is reached. The nonlinear error is, naturally, only present if a nonlinear solver is used. It is relevant only for unstable flows and more critical for large values of $\tau$, when the most unstable modes have enough time to grow and become nonlinear.

\begin{figure}[h!]
	\begin{center}
		\includegraphics[width=1\textwidth]{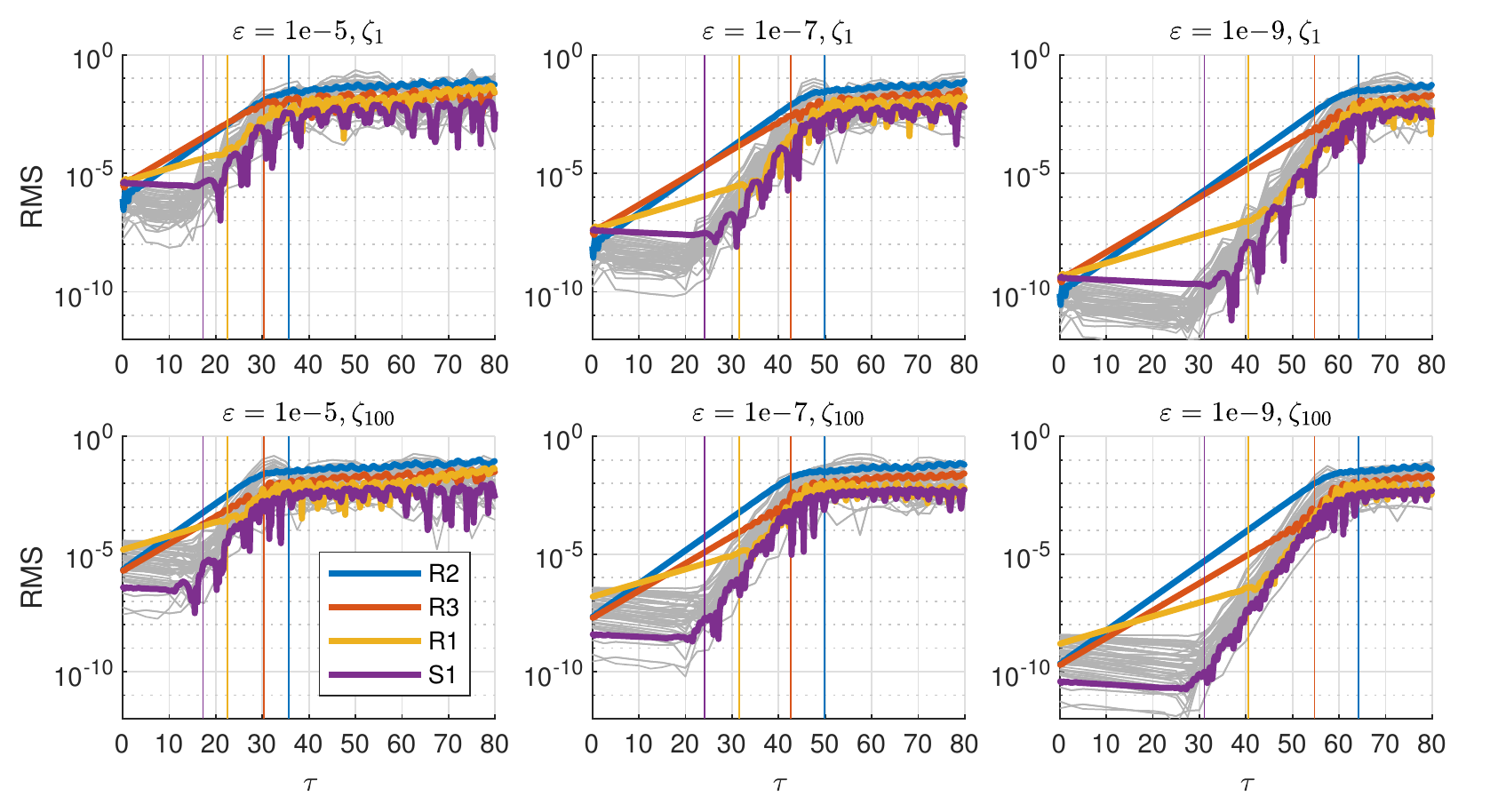}
	\end{center}
	\caption{Evolution of each mode in the flow as it is simulated from different starting conditions. Modes R1 to R3 and S1 are highlighted. Modes up to the 100\textsuperscript{th} least stable are shown as thin gray lines. Vertical lines depict the upper limits of $\tau$, as given by equation~\ref{eq:NLTauBound2}.}
	\label{fig:modesInZetas}
\end{figure}

Figure~\ref{fig:modesInZetas} shows the same analysis as figure~\ref{fig:modesInBaseflow}, but for different starting conditions. In the upper row, we added the initial disturbance of the Arnoldi procedure, $\zeta_1$, to the base flow with magnitudes $\varepsilon=10^{-5}$, $10^{-7}$ and $10^{-9}$. The bottom row uses disturbance $\zeta_{100}$, corresponding to the 100\textsuperscript{th} Arnoldi iteration. Similar results would be obtained if other iterations were chosen. This illustrates the mode evolution within each Arnoldi iteration. In these cases, the nonlinearities surfaced much earlier in the simulation, especially for larger values of $\varepsilon$.

This phenomenon defines an upper bound of $\tau$, which can be estimated as follows. It is required that the mode amplitude be substantially larger than the nonlinear error contamination,

\begin{equation}
	\varepsilon e^{\sigma_i^R \tau} \gg \varepsilon^2 e^{ 2 \sigma_1^R \tau}\ ,
	\label{eq:NLTauBound1}
\end{equation}

\noindent leading to

\begin{equation}
	\tau \ll \frac{\ln(\varepsilon)}{\sigma_i^R-2\sigma_1^R}\ .
	\label{eq:NLTauBound2}
\end{equation}

Using equation~\ref{eq:NLTauBound2}, we computed the $\tau$ bounds associated with each mode, represented by the vertical lines in figure~\ref{fig:modesInZetas}. They offer good estimates of the values of $\tau$ from which the linear projection of each mode shows nonlinear effects. This is also the time from which there is a massive error increment  in figures~\ref{fig:dtConvMode1}~and~\ref{fig:dtConvModeT}. As seen in these figures, the bounds slightly decrease for more stable modes and are not affected by the order of accuracy of the Fréchet derivative. The estimates  captured consistently the effect of $\varepsilon$  and apply for any Krylov iteration. It is important to note that, if there are no unstable modes, these nonlinear effects should reduce with time and the upper bound given by equation~\ref{eq:upperBound1} is likely to be the only constraint for large $\tau$.

As for the lower values of $\tau$, Figures~\ref{fig:dtConvMode1}~and~\ref{fig:dtConvModeT} show a slow increase in error as $\tau$ is reduced. This is associated with the discretization error of the Fréchet derivative approximations, which in turn affects $\sigma_A$. If there is not enough time of flow evolution, the finite differences in the Fréchet derivative, become vulnerable to numerical errors of the flow solver.  

We developed estimates for an optimal $\tau$ as follows. We can include in equation~\ref{eq:error3} the nonlinear error neglected in the previous section, which becomes

\begin{equation}
	E_B = \mathcal{O}((\varepsilon \sqrt{N})^n) + \mathcal{O}(\varepsilon_S/\varepsilon) + \mathcal{O}(\varepsilon e^{ 2 \sigma_1^R \tau})\ .
	\label{eq:errInB}
\end{equation}

From equation~\ref{eq:Bdef}, we have

\begin{equation}
	dA = \frac{1}{\tau} e^{-A \tau} dB\ .
	\label{eq:eA}
\end{equation}

\noindent Substituting $E_B$ from equation \ref{eq:errInB} into $dB$, the norm of the error in $A$ is

\begin{equation}
	E_A  \approx \frac{(\varepsilon \sqrt{N})^n + \varepsilon_S/\varepsilon + \varepsilon e^{ 2 \sigma_1^R \tau}}{\left\Vert \tau e^{A \tau} \right\Vert}\ .
	\label{eq:errInA}
\end{equation}

For small $\tau$, the denominator of equation~\ref{eq:errInA} can  be written as

\begin{equation}
	E_A  \approx \frac{(\varepsilon \sqrt{N})^n + \varepsilon_S/\varepsilon + \varepsilon e^{ 2 \sigma_1^R \tau}}{\tau}\ ,
	\label{eq:errInAsmalltau}
\end{equation}

\noindent which predicts a slow accuracy degradation as $\tau$ reduces. The behavior is consistent with the results shown in figures~\ref{fig:dtConvMode1}~and~\ref{fig:dtConvModeT}. As explained above, the error arises in the Fréchet derivative approximations, but is governed by the accuracy of the flow solver and hence we assign it to errors in the $F$ approximations. The rate of increase of error as $\tau$ reduces is not dependent on either the Fréchet derivative accuracy order or on $\varepsilon$, as suggested by figures 9 and 10 .

If $\tau$ is so large that the leading mode becomes dominant, we approximate$\Vert e^{A \tau} \Vert$ by $e^{\sigma_1^R\tau}$, the leading eigenmode, which yields

\begin{equation}
	E_{A} \approx \frac{(\varepsilon \sqrt{N})^n + \varepsilon_S/\varepsilon + \varepsilon e^{ 2 \sigma_1^R \tau}}{\tau e^{\sigma_1^R\tau}}\ .
	\label{eq:errInA1}
\end{equation}

\begin{figure}[h!]
	\begin{center}
		\includegraphics[width=0.9\textwidth]{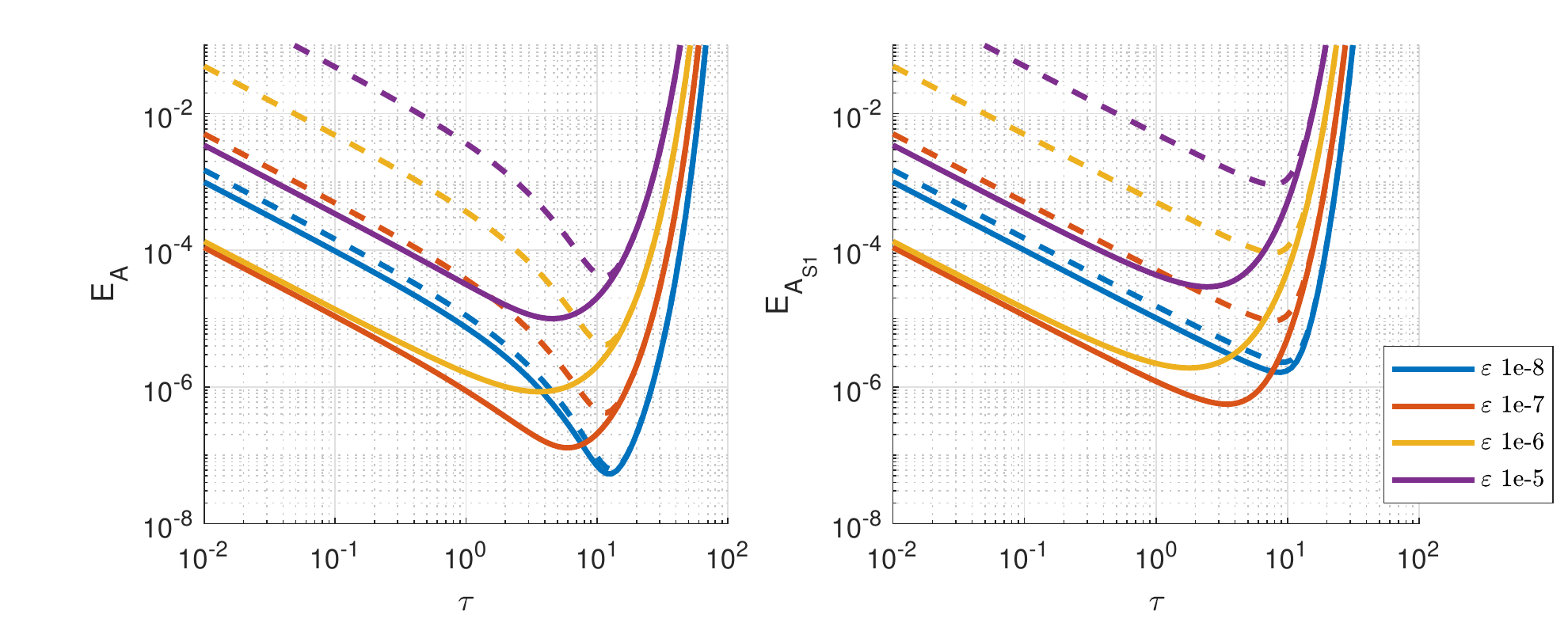}
	\end{center}
	\caption{Estimate of the error in $A \zeta$ as a function of $\tau$ for select values of $\varepsilon$, (Left) using equation~\ref{eq:errInA1} and (Right) using equation~\ref{eq:errInA2} with values corresponding to mode S1. The solid lines represent second order accurate Fréchet derivative approximation and the dashed lines, first order approximation.}
	\label{fig:eaEstimates}
\end{figure}

This equation is plotted in Figure~\ref{fig:eaEstimates}~(Left) for various combinations of $\varepsilon$ and Fréchet derivative order of accuracy approximations. Figures~\ref{fig:dtConvMode1}~and~\ref{fig:dtConvModeT} show the error in the eigenvalues $\sigma$, while equation~\ref{eq:errInA1} and figure~\ref{fig:eaEstimates} relate to the error in $A\zeta$, which is of the same order of the errors in $A$ itself, as $\zeta$ has an unitary norm. However, according to Davis and Moler\cite{Davis1978}, when a small disturbance is added to an invertible matrix, its eigenvalues will be changed by an amount whose upper bound is proportional to the norm of said disturbance. Since matrix $A$ is invertible (as it has no null eigenvalue), this explains why $E_A$ and $E_\sigma$ behave similarly. This also justifies the estimates of optimal $\varepsilon$.

Equation~\ref{eq:errInA1} can provide estimates of the optimal $\tau$ for the dominant mode. However, for less unstable (or more stable) modes, the upper boundary of $\tau$ reduces (see figure~\ref{fig:modesInZetas}) and so would optimal $\tau$. Figures~\ref{fig:modesInBaseflow}~and~\ref{fig:modesInZetas} indicate that each mode behaves independently from the others with regard to nonlinear errors. This observation led us to an ad-hoc equation that offers good estimates of optimal $\tau$ for any given eigenvalue. We simply replace $e^{A\tau}$ by $e^{\sigma_i^R\tau}$ in equation~\ref{eq:errInA} for each mode $i$ yielding

\begin{equation}
	E_{A_i} \approx \frac{(\varepsilon \sqrt{N})^n + \varepsilon_S/\varepsilon + \varepsilon e^{ 2 \sigma_1^R \tau}}{\tau e^{\sigma_i^R\tau}}\ ,
	\label{eq:errInA2}
\end{equation}

\noindent where $E_{A_i}$ is meant to indicate the error in matrix $A$ which would affect accuracy of eigenvalue $i$. For small $\tau$, it reduces to equation~\ref{eq:errInAsmalltau}. Also note that, for large $\tau$ and stable flows, the equation leads to the upper bound expressed by equation~\ref{eq:NLTauBound2}, as presented by Gómez et al.\cite{Gomez2014}. Figure~\ref{fig:eaEstimates}~(Right) shows the value of $E_{A_i}$ for Mode S1, while figure~\ref{fig:eaEstimates}~(Left) can be interpreted as the equivalent for Mode R2. By comparing figure~\ref{fig:eaEstimates} with figures~\ref{fig:dtConvMode1}~and~\ref{fig:dtConvModeT}, we observe that equation \ref{eq:errInA2} adequately accounts for the reduction of the optimum $\tau$ for less unstable modes, see figure~\ref{fig:modesInZetas}. Figure~\ref{fig:eaEstimates} also shows that an overestimation of $\tau_{opt}$ is much more consequential than an underestimation. Therefore, a safe approach is to use a value of $\tau$ somewhat smaller than the optimum; in our study, very good results were produced for $\tau$ about 5 times smaller than the upper limit given by equation~\ref{eq:NLTauBound2}.

\begin{figure}[h!]
	\begin{center}
		\includegraphics[width=0.8\textwidth]{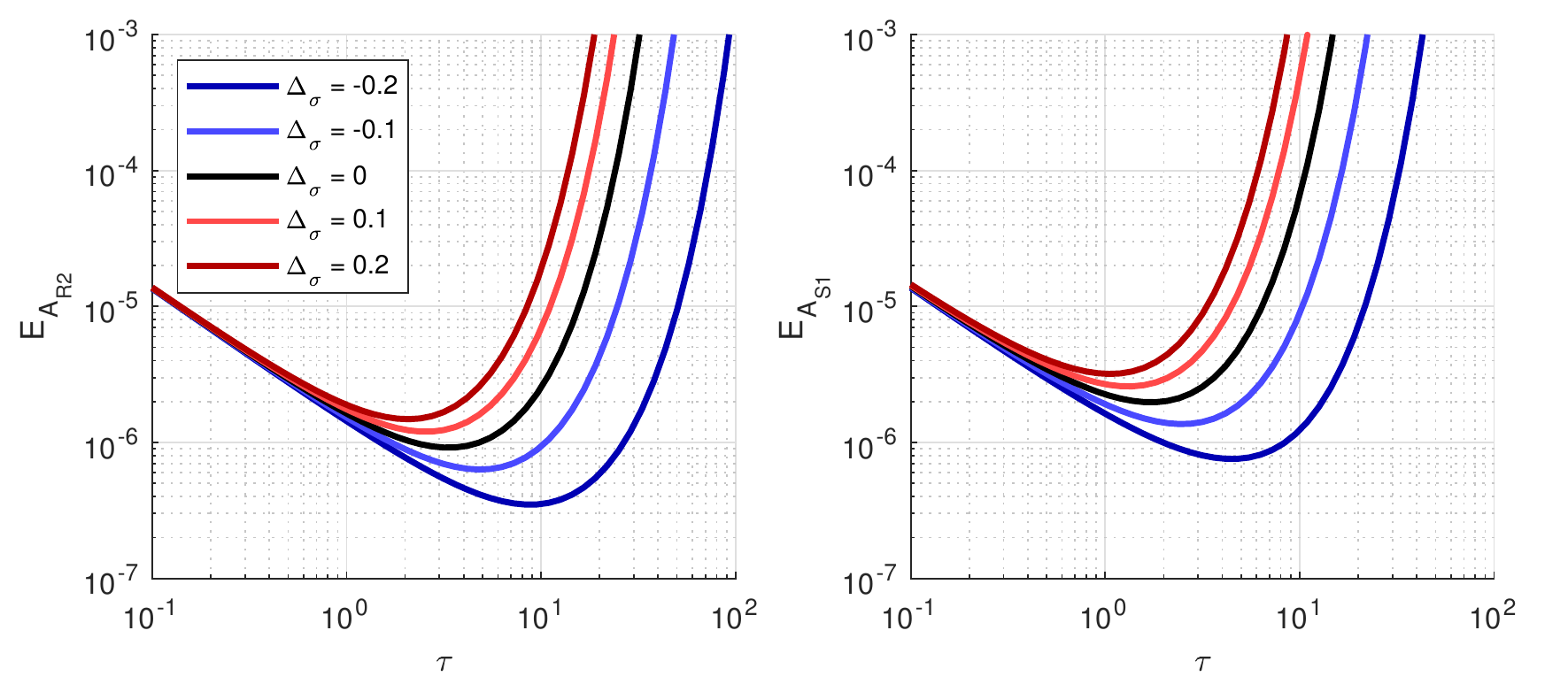}
	\end{center}
	\caption{(Left) Values of $E_A$ for mode R2 when the real part of its eigenvalue is changed. (Center) Values of $E_A$ for mode S1 when the real part of the leading eigenvalue is changed. (Right) Values of $E_A$ for mode S1 when the real part of its eigenvalue is changed.}
	\label{fig:sigmaSensitivity}
\end{figure}

An estimate of the eigenvalues is needed as input for equations~\ref{eq:errInA}~to~\ref{eq:errInA2}, which may not be known a priori. Therefore, it is important to quantify the sensitivity of optimal $\tau$ to the eigenvalues. Figure~\ref{fig:sigmaSensitivity}~(Left) shows the value of $E_A$ for mode R2 as a function of  $\tau$ computed by equation~\ref{eq:errInA2} for a range of $\sigma_{R2}^R$ estimates. $\varepsilon$ was fixed at $10^{-6}$ and the second order accurate Fréchet derivative was used. The actual value, $\sigma_{R2}^R=0.3230$, is indicated by the black line. Each color represents a different value of $\Delta_\sigma$, which was added to $\sigma_{R2}$. Linear stability analysis is most often used close to neutral conditions, where it is more meaningful for the interpretation of the flow behavior. Under these circumstances, the magnitude of the dominant mode growth rate is likely to be less than one. Hence the range of  $\sigma_{R2}^R$ covered is wide and representative. The figure shows that the $\tau_{opt}$ varies roughly within an order of magnitude for both figures. More importantly, the impact of a poor estimate of $\tau_{opt}$ on $E_A$, which is given by the black curve at different values of $\tau$, is very small. As discussed above, it is wise to use $\tau<\tau_{opt}$, and figure~\ref{fig:sigmaSensitivity} shows the impact of an order of magnitude reduction in $\tau$ would be inconsequential to $E_A$. Moreover, and overestimation of $\sigma_{R2}^R$ would also lead to lower and safer $\tau_{opt}$. In any case, a good estimate of the leading eigenvalue can be obtained directly with the residual algorithm \citep{Theofilis2003b}, which can be compared to the first iteration of the Arnoldi algorithm.

Equations~\ref{eq:errInA1}~and~\ref{eq:errInA2} comprise not only the effects of $\tau$, but also that of $\varepsilon$ and of the order of accuracy of the Fréchet derivative approximation investigated in the previous section. It can be used as follows. Following the guidelines above, it is anticipated that,  the error associated with the nonlinear term will be negligible to be on the safe side. Hence, differentiation of either of these equations leads to equation~\ref{eq:error4}. With $\varepsilon_{opt}$ defined, it is possible to obtain the optimal value of $\tau$ from equation~\ref{eq:errInA2} with a rough estimate of the leading eigenvalue and the most stable eigenvalues sought. The equation also allows the evaluation of the effect of increasing the order of accuracy of the Fréchet approximations, of reducing the flow solver noise floor and of increasing the computational grid on the optimal parameters and on the accuracy. 

\begin{figure}[h!]
	\begin{center}
		\includegraphics[width=0.9\textwidth]{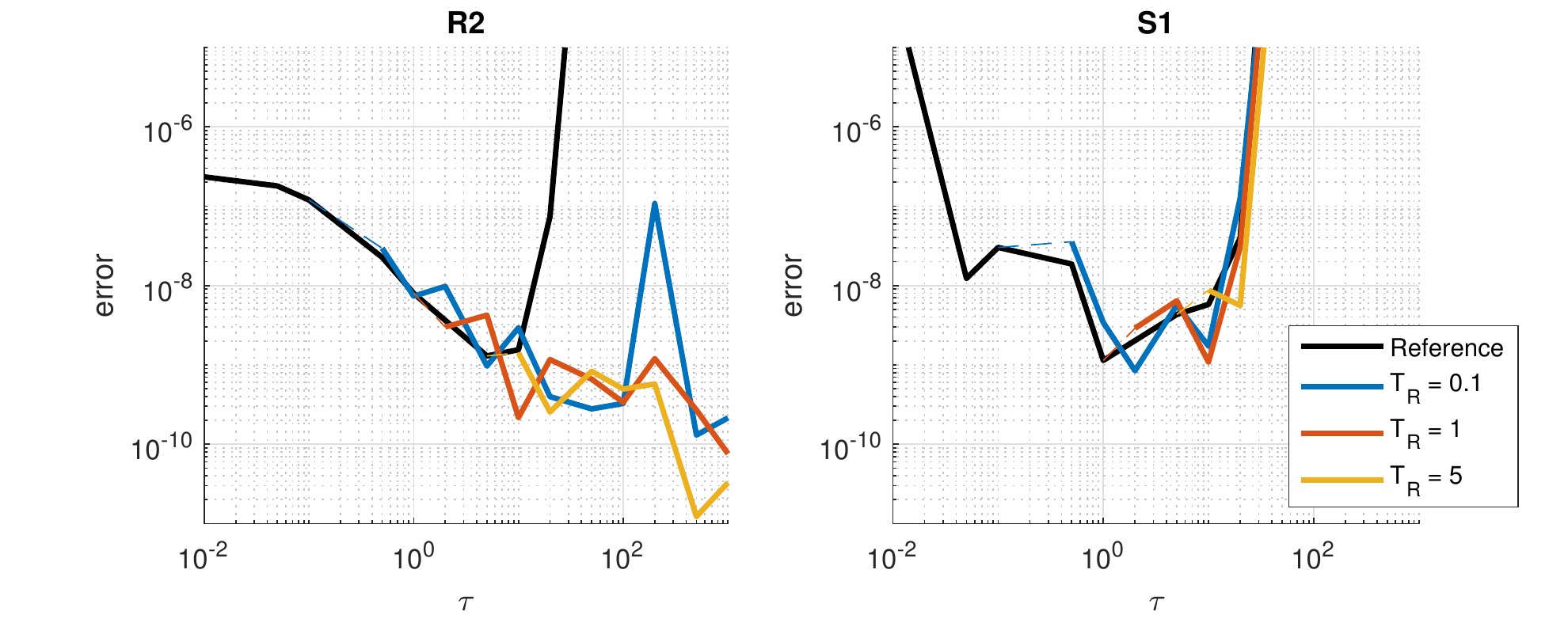}
	\end{center}
	\caption{Difference in the real part of the eigenvalue corresponding to modes R2 (Left) and S1 (Right) as a function of $\tau$ for $\varepsilon = 10^{-7}$ and second order accurate Fréchet derivative. The black line is the reference case, with no renormalization and the colored lines represent different intervals of renormalization.}
	\label{fig:errRenorm}
\end{figure}

It is sometimes argued that a renormalization step can prevent nonlinear effects, as one can increase the maximum value of $\tau$, while keeping the disturbance norm within an acceptable range, away from both nonlinear effects and rounding errors. However, the renormalization of the disturbance would affect all modes equally. Figure~\ref{fig:errRenorm} shows the measured error for modes R2 nd S1 for various values of $\tau$, with $\varepsilon = 10^{-7}$ and second order accurate Fréchet derivative and compares the regular case to cases that were renormalized at different intervals $T_R$ and the reference case. A somewhat irregular improvement in accuracy can be obtained for the dominant mode. For the other modes, the error and the optimal $\tau$ are remarkably similar to the reference case. In fact, each iteration of the Arnoldi algorithm involves a renormalization; however, it is accompanied by an orthogonalization which reshuffles the nonlinear content preventing the contamination. Therefore, renormalization may help improving precision for the leading mode.

\section{Krylov span dimension and optimization of the total computational time}
\label{sec:runTime}

Following Goldhirsch et al.\cite{Goldhirsch1987}, the origin of the lower boundary for $\tau M$ (equation~\ref{eq:lowerBound}) is the error caused by truncating the eigenspectrum to the first $M$ modes, which is

\begin{equation}
	Error \propto \abs{e^{\left( \sigma_{M+1} - \sigma_{1} \right) \tau M}} \ .
	\label{eq:truncationError}
\end{equation}

\noindent This reflects the observation that, in case the growth rates are too close to each other, the total integration time ($\tau M$) must be long enough for a mode to be properly identified, hence the boundary. At a glance, it may seem that any combination of $\tau$ and $M$ that gives the same total filtering time (as they refer to $\tau M$) would produce the same level of accuracy. Nevertheless, increasing $M$ has the additional benefit of moving $\sigma_{M+1}$ further to the left of the spectrum, hence further reducing this type of error. This suggests that increasing $M$ instead of $\tau$ is more cost effective. However, as discussed in section~\ref{sec:compCost}, for large values of $M$ the cost tends to increase nonlinearly with $M$.

\begin{figure}[h!]
	\begin{center}
		\includegraphics[width=0.9\textwidth]{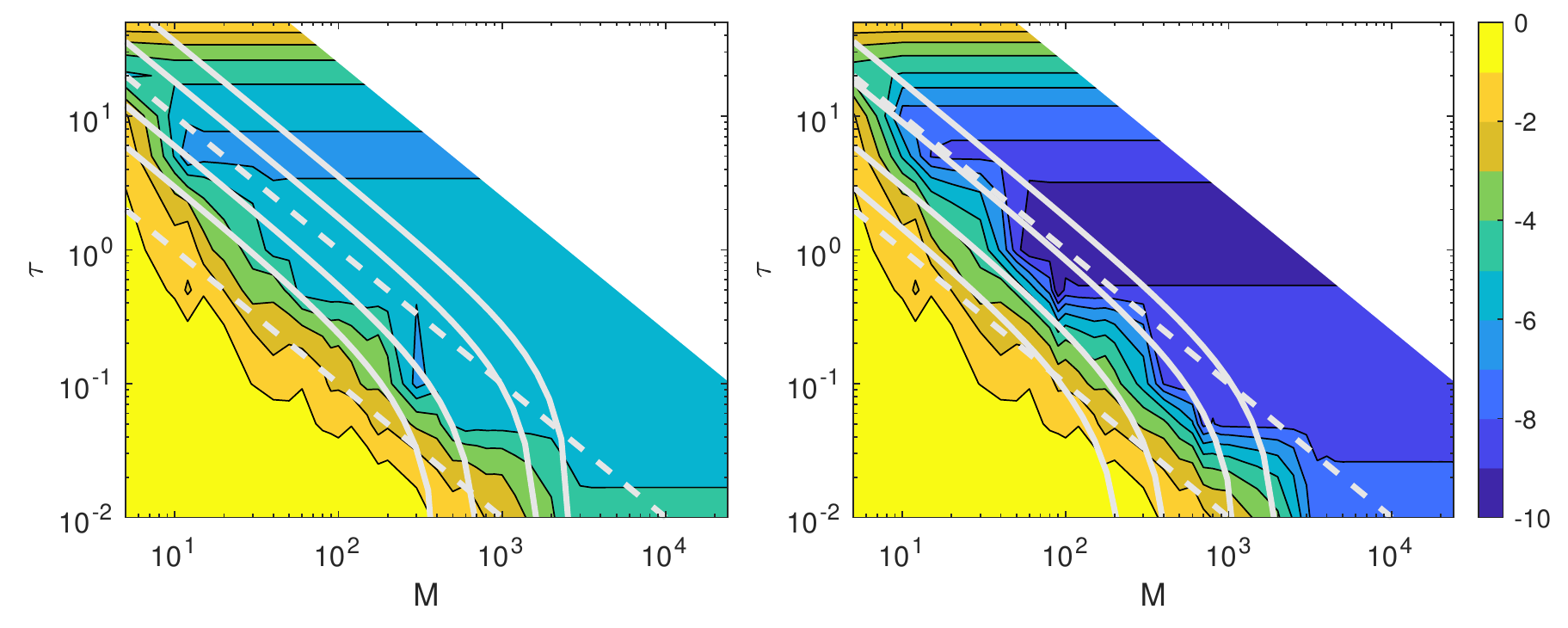}
	\end{center}
	\caption{Isocontours of log$_{10}$(Error) of the real part of the R2 mode, using $\varepsilon=10^{-6}$. The full gray lines represent a constant computational cost of 5 minutes, 10 minutes, 30 minutes and 1 hour from left to right. The dashed gray lines indicate constant $t=\tau M$. (Left)~First order accurate Fréchet derivative. (Right)~Second order accurate Fréchet derivative.}
	\label{fig:map_m1e6}
\end{figure}

\begin{figure}[h!]
	\begin{center}
		\includegraphics[width=0.9\textwidth]{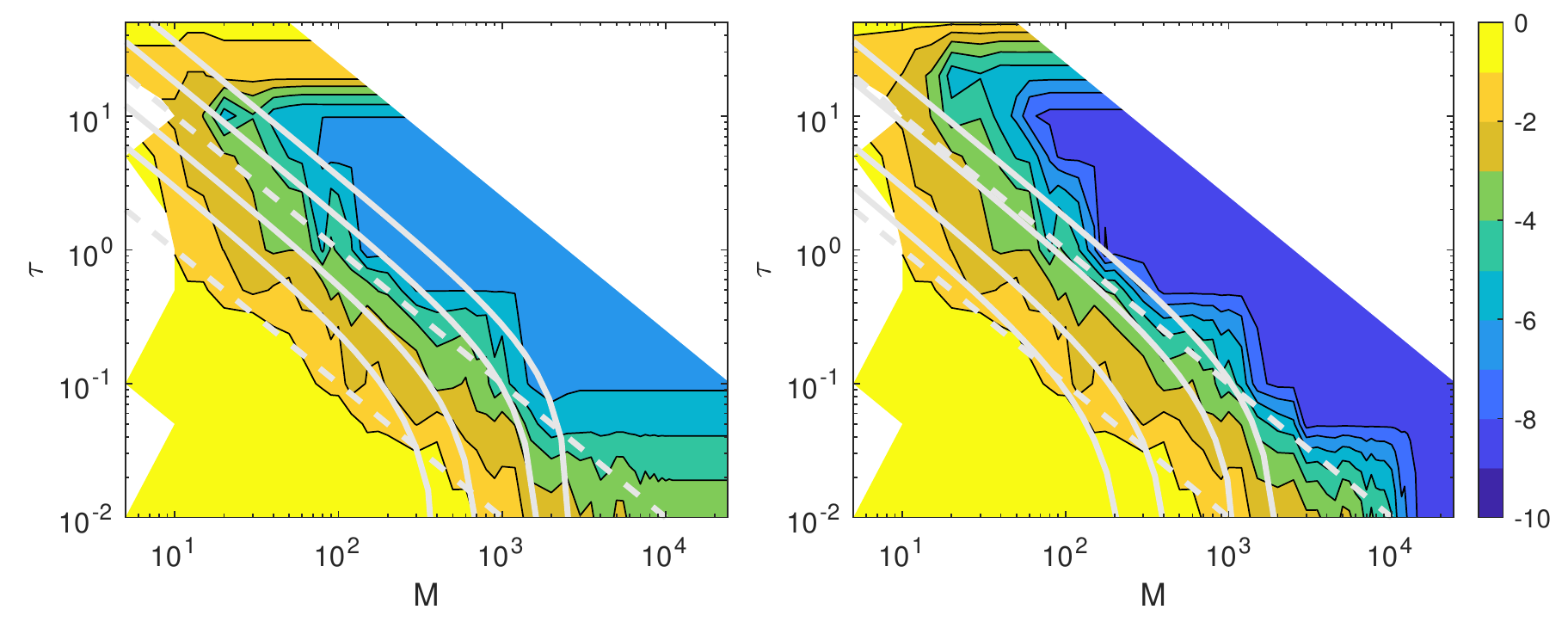}
	\end{center}
	\caption{Isocontours of log$_{10}$(Error) of the real part of the S1 mode, using $\varepsilon=10^{-6}$. The full gray lines represent a constant computational cost of 5 minutes, 10 minutes, 30 minutes and 1 hour from left to right. The dashed gray lines indicate constant $t=\tau M$. (Left)~First order accurate Fréchet derivative. (Right)~Second order accurate Fréchet derivative.}
	\label{fig:map_mTe6}
\end{figure}

Hence there remains the question of the optimal balance between $\tau$ and $M$ that allows the required modes to be retrieved with the desired accuracy and lowest $C$. With this goal, we mapped the error for various combinations of $\tau$ and $M$, for $\varepsilon=10^{-6}$ and for first and second-order accurate Fréchet derivatives. Figures~\ref{fig:map_m1e6}~and~\ref{fig:map_mTe6} give the results for modes R2 and S1, respectively. The colored contours indicate iso-levels of error relative to the reference case. The white area for large $\tau$ and $M$ was not covered in the study, while in the white areas for small $M$, the algorithm did not find the sought mode. The solid light gray lines are isocontours of computational cost, as computed by equation~\ref{eq:computationalCost} with the coefficients from table~\ref{tab:computationalCostCoeff}. From left to right, they represent 5 minutes, 10 minutes, 30 minutes and 1 hour of wall-time, excluding start-up time which is independent of $\tau$ and $M$. The dashed gray lines indicate constant values of $t=\tau M$. Note that, contrary to the previous two sections, this section also addresses the issue of computational time required for a given accuracy and as such the numerical results are dependent on the speed of the flow solver, the efficiency of the eigenvalue problem solver and the computational resources. Nevertheless, the conclusions still serve as general guidelines for improving solution efficiency.

Consistent with Goldhirsch et al.\cite{Goldhirsch1987}, in both figures~\ref{fig:map_m1e6}~and~\ref{fig:map_mTe6}, the error is larger if the product $\tau M$ is small. The bounds for $\tau$ discussed in the previous section are also observable in the figures. At large $\tau$ there are the nonlinear effects which rapidly reduce accuracy. At small $\tau$ there is a gradual reduction in accuracy at large $M$.

Both figures illustrate interesting aspects of the trade-off between $\tau$ and $M$. For $M<20$, iso-accuracy lines approach the vertical, indicating that, in this region, increasing $M$ is more beneficial than increasing $\tau$. This may be a manifestation of the extra benefit of increasing $M$ discussed in connection with equation~\ref{eq:truncationError}. However, at least for the unstable mode, figure~\ref{fig:dtConvMode1} suggests this may also be a manifestation of errors associated with nonlinear effects caused by large $\tau$. Away from this region, the extra benefit of increasing $M$ does not appear to be relevant.

In a large central portion of the figures, the curves of iso-cost are roughly parallel to the constant $\tau M$ lines. This corresponds to the region were the cost of running the flow solver dominates the computations, see figure~\ref{fig:compCost}. At large $M$, the cost increases massively, becoming almost independent of $\tau$. It must also be considered that this argument disregards the fact that increasing $M$ has the added benefit of retrieving more modes; that is, we assumed a given set of eigenvalues is desired.

The results indicate that the most cost-effective approach is to use a total integration time $\tau M$ with the largest $\tau$ possible that is still below the upper bounds established, which, from section~\ref{sec:tau}, tends to provide better accuracy. Under these circumstances the cost is essentially that of running the flow solver. We recall that the cost was measured by wall-time, which, in our solver, is essentially proportional to CPU time; hence, similar conclusions would have been reached if either were the measure of cost. Therefore, optimal $\tau$ does not depend on $M$, which once more removes one dimension of the parametric search.  After establishing $\varepsilon$ and $\tau$ from the equations of the previous sections, we should run the Arnoldi algorithm and monitor the solution for increasing $M$ until the desired accuracy or minimum error bound is reached. As opposed to $\tau$ and $\varepsilon$, $M$ does not need to be decided in advance and its effect on the solution can monitored as the algorithm runs. Our study was restricted to 2D simulations. For 3D simulations, the wall-time or CPU-time associated with the flow integration would be substantially longer and the conclusion would still hold.

Restarted Arnoldi methods substantially reduce the cost associated with the Gram-Schmidt orthonormalization, but, as long as the cost is dominated by the flow solver,this strategy would not be beneficial. This is the situation in our test case, and the reason restarted methods were not considered here as it would not affect the optimal parameters. For 3D simulations the CPU-time associated with the flow solver tends be even longer and even less likely to benefit from restarted algorithms. Overall, our results suggest that a wide range of applications would not benefit from restarted algorithms.

\begin{figure}[h!]
	\begin{center}
		\includegraphics[width=1\textwidth]{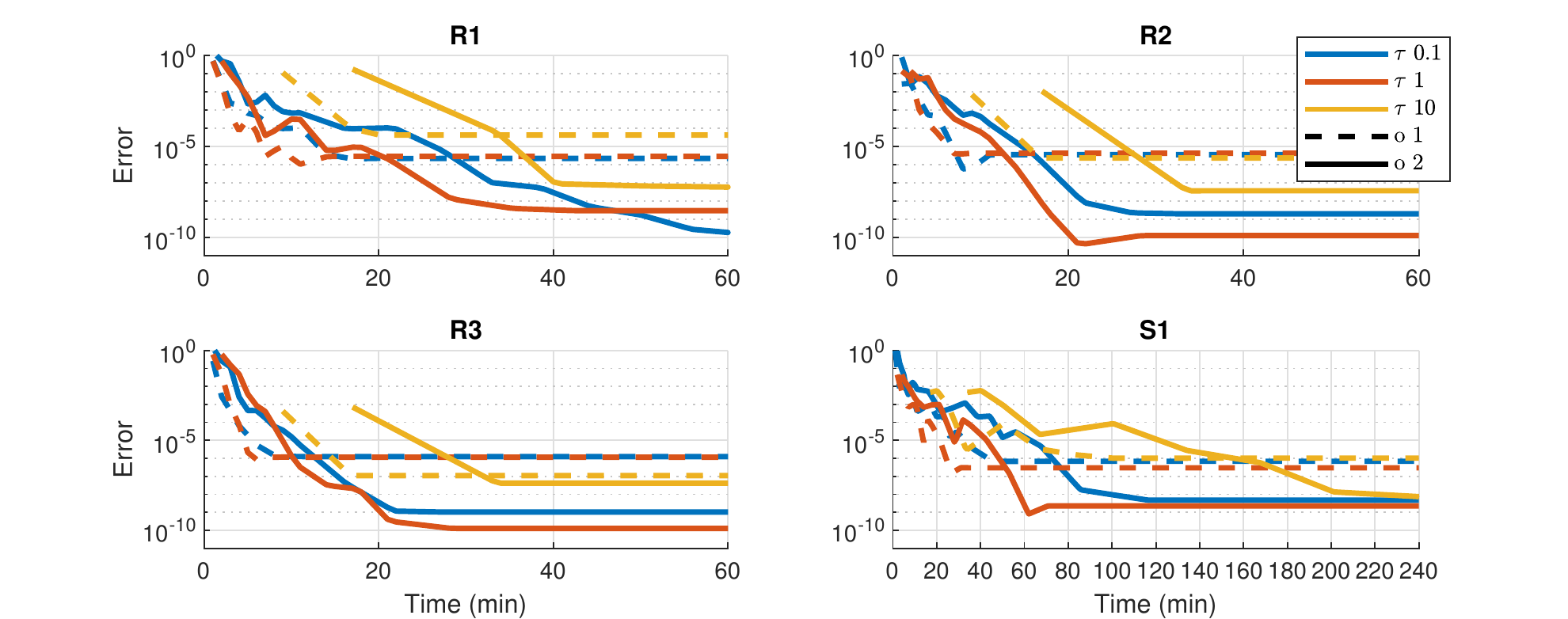}
	\end{center}
	\caption{Error of the first four eigenvalues as computational time is increased. $\tau=0.1,1,10$, $\varepsilon=10^{-6}$. Note that the time scale for mode S1 is different than that of modes R1 to R3.}
	\label{fig:errVsTime}
\end{figure}

The similarity between both frames in each of the figures~\ref{fig:map_m1e6}~and~\ref{fig:map_mTe6} for contour levels down to $10^{-5}$ is remarkable. This indicates that, for these error levels, the truncation order of the Fréchet derivative does not play a role, as discussed in section~\ref{sec:epsilon}. There is an exception for very small $\tau$, for which there is an effect of the truncation error on levels above $10^{-5}$, which is consistent with section~\ref{sec:tau}. Hence, under these circumstances, the first-order is the most efficient approach as it requires shorter computation times for the same accuracy. On the other hand, first-order accuracy could not reach an error level below $10^{-6}$, while second-order accuracy consistently provides errors of $10^{-8}$. This is more clearly illustrated and extended for other eigenvalues in Figure~\ref{fig:errVsTime} which shows the eigenvalue convergence history for first and second-order accurate Fréchet approximations, and for different integration times.

Overall, the results show that, for this problem, if the target error level is about $10^{-5}$, first-order is the better choice, whereas, if errors below $10^{-6}$ are required, at least second-order accurate Fréchet is needed. It should be added that $\varepsilon=10^{-6}$ is optimal for second order accurate Fréchet derivative approximations, and suboptimal for first order; therefore, it is possible that slightly more accurate results would still be possible with first order. In the choice of the target accuracy, one must also consider that, for open flow applications, for example, details of the far field boundary conditions can easily affect the computation of leading eigenvalues in the range of $10^{-5}$ or above \citep{Ohmichi2016}.

\section{Rate of convergence of eigenfunctions and eigenvalues}
\label{sec:eigfunctions}

In this section, we seek to evaluate how the accuracy of the eigenfunction estimates is affected by the parameters chosen, as well as how it affects the eigenvalue convergence rate. Figure~\ref{fig:errVsTime} shows that the more unstable modes reach minimum error progressively faster. In fact, for mode S1, the error starts decreasing consistently only after all unstable modes have approached their minimum error values. This is associated with eigenfunction representation in the Krylov subspace, as we will discuss. Hence, it is important to investigate the accuracy of the eigenfunction estimates.

\begin{figure}[h!]
	\begin{center}
		\includegraphics[width=0.8\textwidth]{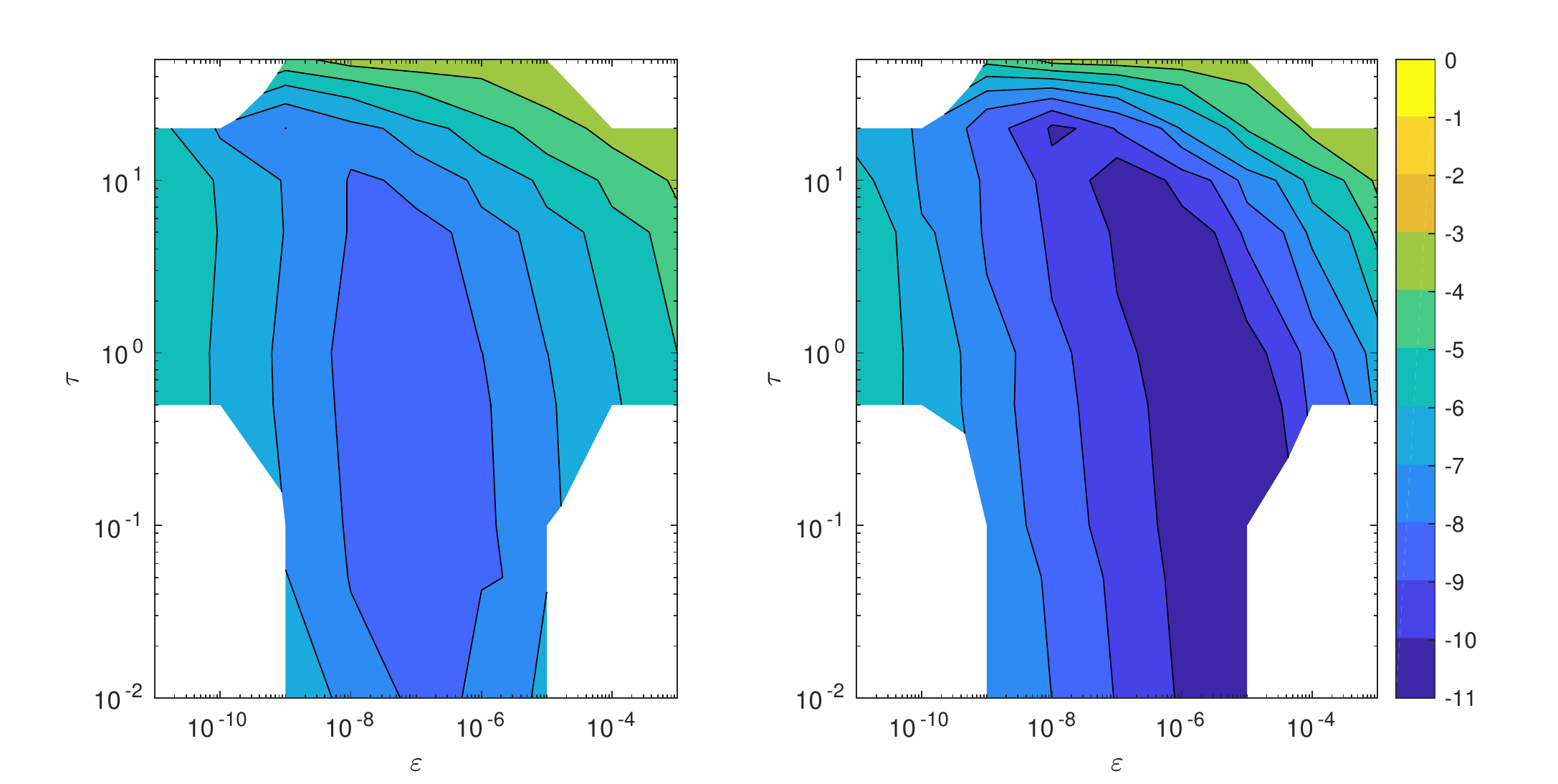}
	\end{center}
	\caption{Isocontours of log$_{10}$(Error) of the eigenfunction of R2 for (Left) first-order accurate Fréchet derivative and (Right) second-order derivative.}
	\label{fig:eigFuncErr1}
\end{figure}

\begin{figure}[h!]
	\begin{center}
		\includegraphics[width=0.8\textwidth]{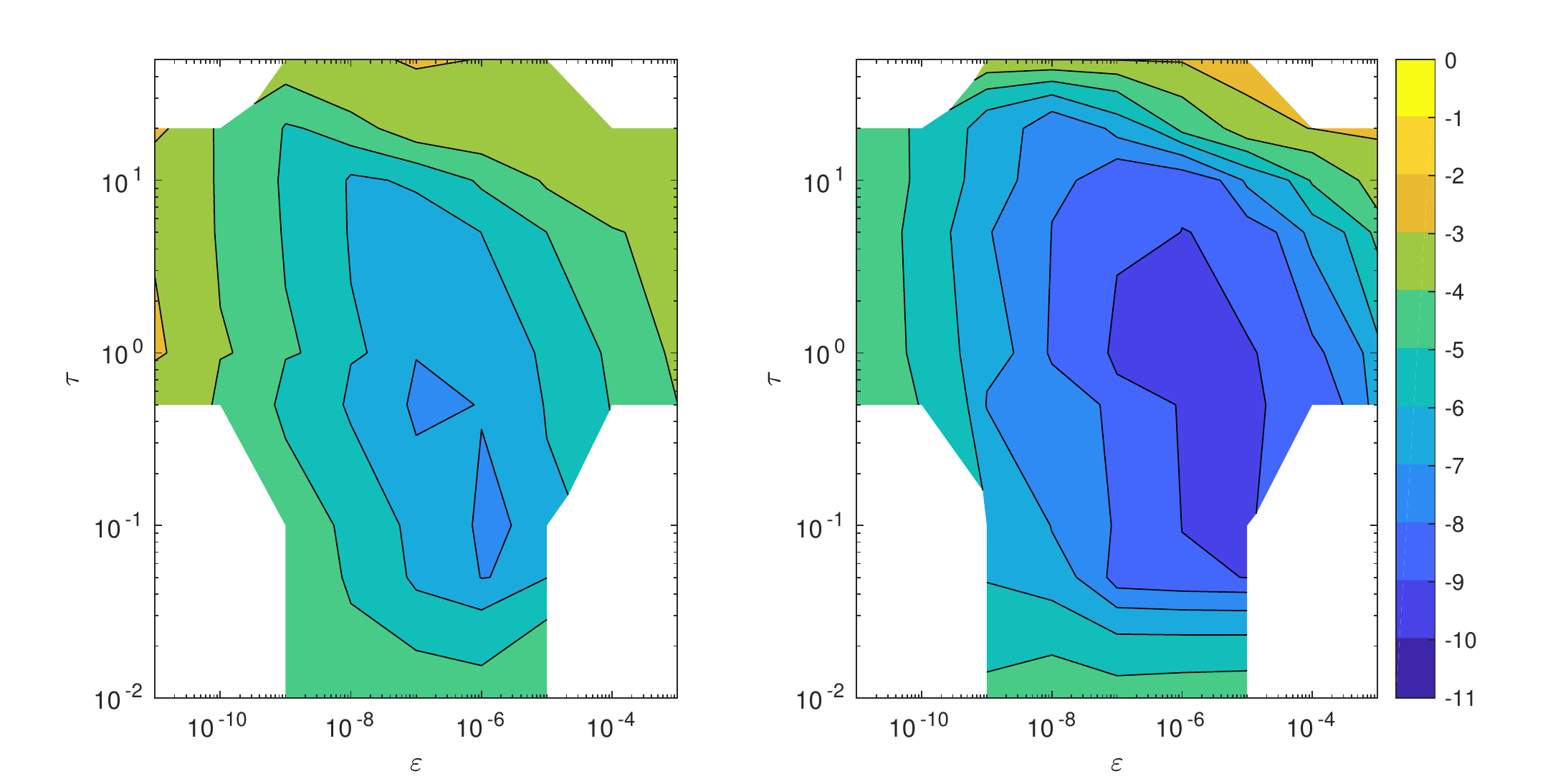}
	\end{center}
	\caption{Isocontours of log$_{10}$(Error) of the eigenfunction of S1 for (Left) first-order accurate Fréchet derivative and (Right) second-order derivative.}
	\label{fig:eigFuncErrT}
\end{figure}

Figures~\ref{fig:eigFuncErr1}~and~\ref{fig:eigFuncErrT} show the error of the eigenfunctions retrieved for modes R2 and S1, respectively. This error is computed as the mean absolute difference between an eigenfunction and its reference, both eigenfunctions have unitary norm $\ell_2$. Using the absolute value of the eigenfunction, our results are unaffected by differences in the eigenfunction phase. The white areas close to the corners of the figures were not included in our parametric sweep.

Consistently, the parameters leading to most accurate eigenvalue results worked for the eigenfunctions as well. Looking at mode R2, the eigenfunction reaches an accuracy of up to $10^{-9}$ with a first-order accurate Fréchet derivative and up to $10^{-11}$ with a second-order one. Mode S1 reached an accuracy $10^{-7}$ with a first-order accurate derivative, while the second-order one achieved $10^{-9}$.

In some ways, these pictures summarize results obtained for the eigenvalues too. At large $\tau$ there are nonlinear effects that affect the accuracy. For small $\tau$ and small $\varepsilon$ the computation of $F$ becomes vulnerable to the discretization error. For large $\varepsilon$, there are both nonlinear and Fréchet derivative truncation errors, while, at low $\varepsilon$, there are discretization errors in $F$. Because of these effects, the region of maximum accuracy lies at the center.  For second order accurate Fréchet derivatives, the truncation error reduces more rapidly, such that for this case the optimal region moves to the right as compared to the first order accurate one.

\begin{figure}[h!]
	\begin{center}
		\includegraphics[width=0.6\textwidth]{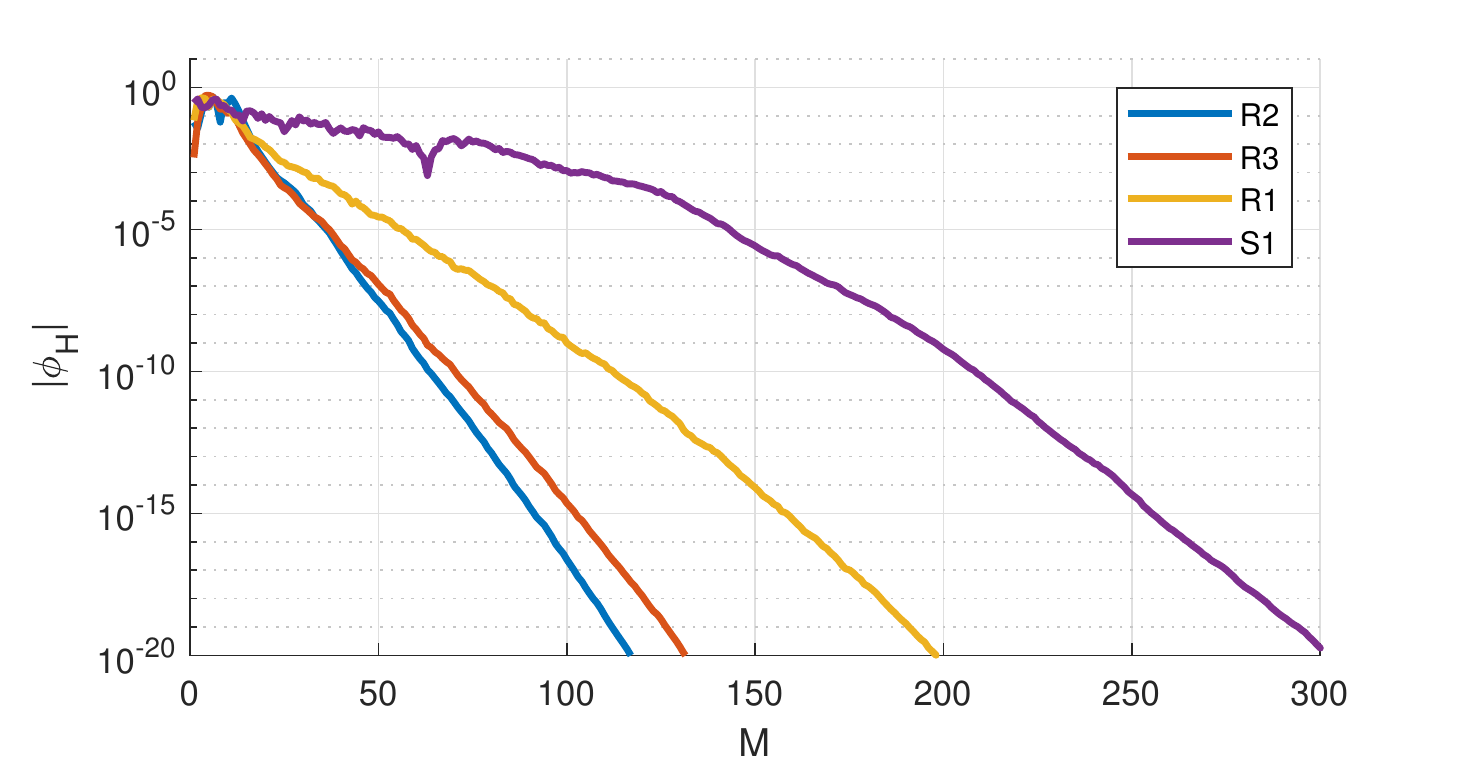}
	\end{center}
	\caption{Weight of each disturbance $\zeta_i$ in the eigenfunctions of modes R1, R2, R3 and S1.}
	\label{fig:eigFuncConv}
\end{figure}

In the time-stepping method discussed, the eigenfunctions of each flow mode are retrieved as a linear combination of $\zeta_i$, that are the initial disturbances in each iteration of the Arnoldi procedure. The weight of each $\zeta_i$ is given by the eigenvectors of matrix $H$, as per step 12 of Algorithm~\ref{algorithm}. By examining the eigenvectors $\phi_H$, we can infer how much each iteration contributes to the convergence of a mode. Figure~\ref{fig:eigFuncConv} shows $\abs{\phi_H}$ as a function of M for R1, R2, R3 and S1 in our reference case. The most unstable modes converge faster. After a mode reaches convergence and therefore can be accurately filtered out of the flow by the Gram-Schmidt orthonormalization, the convergence rate of the next one is accelerated. This is because the converged mode can be accurately removed from the next disturbance by the Gram-Schmidt algorithm, which hence concentrates on filtering of the next dominant mode. It must also be added that modes which are almost orthogonal are expected to have smaller effects on each other. This is possibly an explanation for the faster convergence of mode R3 relative to R1. The convergence of the eigenfunction closely follows, and also explains, the convergence of eigenvalues in figure~\ref{fig:errVsTime}.

\section{Final remarks and conclusion}

We sought to establish guidelines for choosing computational parameters when performing a time-stepping, Jacobian-free global instability analysis of a flow. This procedure requires a solver that produces a flow solution, $F$. The computational parameters considered were disturbance magnitude ($\varepsilon$), order of accuracy of the Fréchet derivative approximation ($n$), integration time ($\tau$) and dimension of the Krylov span ($M$). The analysis was carried out with codes developed in-house and used a two-dimensional compressible open cavity flow as an example; however, these guidelines should be valid for other types of flows and solvers, including three-dimensional scenarios. The following sources of error may exist in this problem: (1) truncation error associated with the Fréchet derivative approximations; (2) discretization errors in the computation of $F$, which can be truncation or round-off; (3) errors associated with the nonlinear terms of the Navier-Stokes equations; and (4) errors associated with the Krylov subspace approximation. Parameter $n$ affects the truncation error (1). Parameter $\varepsilon$ affects the truncation (1), discretization (2) and the nonlinear (3) errors. Parameter $M$ affects the Krylov subspace approximation (4) only, while parameter $\tau$ affects all errors.

In principle, the choice of $\varepsilon$, $\tau$ and $M$ constitutes a search in a three-dimensional space. However, our analysis demonstrated that optimal $\tau$ must correspond to effectively negligible nonlinear terms. This led to the optimal $\varepsilon$ being independent of $\tau$ and $M$, which reduces one dimension in the parametric search. The analysis also showed that optimal $\tau$ is independent on $M$ and leads to the most cost-effective choice of $\tau M$, which reduces another dimension in the parametric search.  Hence, as a guideline, one must first determine $\varepsilon_{opt}$, next $\tau_{opt}$ and finally $M$. $M$ does not need to be known a priori and can be monitored along the Arnoldi iterations at no cost.

It was further found that optimal values of $\varepsilon$ depend on a balance between truncation error (1) and discretization error (2) and can be estimated a priori by equation \ref{eq:error4}. Equation~\ref{eq:error3} can  be used to evaluate the benefit of increasing the accuracy order of the Fréchet derivative approximations, the consequences of using less accurate flow solvers, and the care that must be taken when large grids are employed. The results show that increasing the accuracy order of the Fréchet derivative has little impact on the results of less accurate codes with large grids. This is a timely aspect in view that, recently, second order accurate codes are being used for time-stepping instability analysis \citep{Gomez2014,Loiseau2019}.

Our analysis shows that optimal values of $\tau$ can be established primarily based on its impact on the error associated with computing $F$ and the Fréchet derivative, disregarding its impact on the Krylov span approximation. Under these circumstances, optimal $\tau$ depends on a complicated balance between the truncation error (1) and the discretization error (2), which both grow as $\tau$ reduces, and the nonlinear term error (3), which grows as $\tau$ increases. As such, optimal $\tau$ depends on $\varepsilon$, $\varepsilon_S$, $n$, on the mesh size, and on the growth rates of the modes involved. It is possible to estimate optimal $\tau$ by finding the minimum of equation~\ref{eq:errInA2}. The magnitudes of the eigenvalues are required, but a guess would be enough. Figure~\ref{fig:eaEstimates} shows that the error increases massively for $\tau>\tau_{opt}$ and much less so for $\tau<\tau_{opt}$, hence, to be on the safe side, it is advisable to use values of $\tau$ somewhat smaller than $\tau_{opt}$, perhaps a factor of 5. It is important to note that the discussion focused on a situation where a nonlinear solver is used for an unstable flow, which is common. If a linear solver is used, upper bounds for $\tau$ apply only for stable modes and can be obtained from equation \ref{eq:upperBound1} \citep{Gomez2014}, which is also embedded in equation~\ref{eq:errInA2}.

In our study, with optimal parameters, second order accurate Fréchet derivative approximations consistently provided eigenvalues within an accuracy of $10^{-8}$. First order ones consistently offered eigenvalue accuracy of $10^{-5}$, even with slightly suboptimal parameters. If this reduced accuracy is sufficient, first order can deliver it at about half the cost of the second order one, as it needs only one call to the flow solver per Arnoldi iteration. For open flows, small variations in the numerical boundary conditions used to represent an infinite domain can easily affect the fourth or even the third decimal place of the eigenvalue computation \citep{Ohmichi2016}. Hence, with the optimal values of $\varepsilon$ and $\tau$, it is likely that, in many applications, first order accurate Fréchet derivative approximations would be the most cost-effective choice.

Parameter $M$ affects only the accuracy of the Krylov span approximations, in the form of its product with $\tau$. In fact, our tests demonstrate that the accuracy increases with $\tau M$, as explained by Goldhirsch et al.\cite{Goldhirsch1987}. $M$, however, has a more complicated impact on computation time because the Gram-Schmidt procedure scales with $M^2$ and the eigenvalue problem with around $M^3$, thus it is interesting to reduce $M$. Hence using optimal $\tau$ also leads to the most cost-effective approach. Under these circumstances, the computational effort is used almost entirely in the flow solver, and scales linearly with $M$. Hence, optimizations in the orthogonalization or in the eigenvalue solver algorithm have almost no impact. In contrast to $\varepsilon$ and $\tau$, there is not an optimal $M$ for maximum accuracy. As $M$ increases, the error in the eigenvalues eventually reduces to a noise floor associated with the other errors. As the Arnoldi algorithm evolves, the accuracy of the eigenvalues can be monitored until the target level or the noise floor is achieved. Contrary to $\varepsilon$ and $\tau$, the desired $M$ does not need to be known a priori. If three-dimensional flows are investigated, the cost associated with the flow solver would be even larger and the conclusions would still hold. In our analysis the wall-time was used as a measure of cost. In many applications, CPU-time could be a better measure; however, given that at optimal conditions the cost is dominated by the flow solver, using CPU-time as a measure of cost would lead to the same guidelines. It is important to mention that in situations such as our test 2D case where the computational cost is dominated by the flow solver, the used of restarted Arnold methods does not improve efficiency, and for that reason it was not studied here. In 3D global instability analysis, restarted methods are even less likely to be advantageous.

Beyond these guidelines, we investigated the rate of convergence of the eigenvalues and observed that an eigenvalue initiates a rapid convergence only after the eigenfunction of some previous eigenvalues have converged. This is associated with the level of orthogonality between these eigenfunctions. Hence, eigenvalue convergence requires eigenfunction convergence and the current guidelines work for both.

\bmhead{Acknowledgments}

The authors would like to thank the São Paulo Research Foundation (FAPESP/Brazil), for grants 2018/04584-0, 2017/23622-8 and 2019/15366-7; the National Council for Scientific and Technological Development (CNPq/Brazil) for grants 134722/2016-7 and 307956/2019-9; the US Air Force Office of Scientific Research (AFOSR) for grant FA9550-18-1-0112, managed by Dr. Geoff Andersen from SOARD; the University of Liverpool for the access to the Barkla cluster, provided by Prof. Vassilios Theofilis; the Center for Mathematical Sciences Applied to Industry (CeMEAI) funded by the São Paulo Research Foundation (FAPESP/Brazil), grant 2013/07375-0, for access to the Euler cluster, provided by Prof. José Alberto Cuminato; and the National Laboratory for Scientific Computing (LNCC/MCTI, Brazil) for providing HPC resources of the SDumont supercomputer. We also would like to thank Prof. Sávio Brochini Rodrigues, of the Federal University of São Carlos, for the valuable discussions and inputs during the development of this work.


\bibliography{library}


\end{document}